\documentclass[aps,prb,twocolumn,superscriptaddress,nofootinbib,10pt]{revtex4-2}
\usepackage{amsfonts}
\usepackage{amsmath}
\usepackage{amssymb}
\usepackage{amsthm}
\usepackage{mathrsfs}
\usepackage{latexsym}
\usepackage[table, dvipsnames]{xcolor} 
\usepackage{multirow} 
\usepackage{dirtytalk} 
\usepackage{hyperref}
\hypersetup{colorlinks=true, linkcolor=blue, citecolor=red, filecolor=blue, menucolor = green, runcolor = cyan, urlcolor=blue}
\usepackage{subfig} 
\usepackage{enumerate} 
\usepackage{geometry} 
\usepackage{graphicx} 
\usepackage{tikz} 
\usepackage{orcidlink}
\usetikzlibrary{quantikz}
\usetikzlibrary{decorations.pathreplacing}
\usetikzlibrary{3d,calc}
\usetikzlibrary{shapes.geometric}
\usepackage{tcolorbox} 
\usepackage{booktabs} 
\usepackage{lipsum}
\usepackage{tikzpeople}
\usepackage{float} 
\usepackage[normalem]{ulem}
\usepackage{comment}
\usepackage{algpseudocode}
\usepackage{algorithm}
\usepackage{adjustbox}
\usepackage{framed}
\usepackage{ulem}
\usepackage{bm}
\usepackage{ragged2e}

\newcommand{\R}{\mathbb{R}}
\newcommand{\C}{\mathbb{C}}

\newcommand{\N}{\mathbb{N}}

\newcommand{\F}{\mathbb{F}}

\begin{document}

\title{Maritime object classification with SAR imagery using quantum kernel methods}

\author{John Tanner\orcidlink{0000-0002-7458-8295}}
\email{john.tanner@uwa.edu.au}
\affiliation{Centre for Quantum Information, Simulation and Algorithms, The University of Western Australia, 35 Stirling Hwy, Crawley WA, 6009, Australia}
\affiliation{School of Physics, Mathematics and Computing, The University of Western Australia, 35 Stirling Hwy, Crawley WA, 6009, Australia}

\author{Nicholas Davies\orcidlink{0009-0001-7017-8926}}
\affiliation{Centre for Quantum Information, Simulation and Algorithms, The University of Western Australia, 35 Stirling Hwy, Crawley WA, 6009, Australia}
\affiliation{School of Physics, Mathematics and Computing, The University of Western Australia, 35 Stirling Hwy, Crawley WA, 6009, Australia}

\author{Pascal Jahan Elahi\orcidlink{0000-0002-6154-7224}}
\affiliation{Pawsey Supercomputing Centre, 1 Bryce Avenue, Kensington WA, 6151, Australia}
\affiliation{Centre for Quantum Information, Simulation and Algorithms, The University of Western Australia, 35 Stirling Hwy, Crawley WA, 6009, Australia}
\affiliation{School of Physics, Mathematics and Computing, The University of Western Australia, 35 Stirling Hwy, Crawley WA, 6009, Australia}

\author{Casey R.~Myers\orcidlink{0000-0002-8838-7523}}
\affiliation{School of Physics, Mathematics and Computing, The University of Western Australia, 35 Stirling Hwy, Crawley WA, 6009, Australia}
\affiliation{Pawsey Supercomputing Centre, 1 Bryce Avenue, Kensington WA, 6151, Australia}

\author{Du Huynh\orcidlink{0000-0003-3080-9655}}
\affiliation{Centre for Quantum Information, Simulation and Algorithms, The University of Western Australia, 35 Stirling Hwy, Crawley WA, 6009, Australia}
\affiliation{School of Physics, Mathematics and Computing, The University of Western Australia, 35 Stirling Hwy, Crawley WA, 6009, Australia}

\author{Wei Liu\orcidlink{0000-0002-7409-0948}}
\affiliation{Centre for Quantum Information, Simulation and Algorithms, The University of Western Australia, 35 Stirling Hwy, Crawley WA, 6009, Australia}
\affiliation{School of Physics, Mathematics and Computing, The University of Western Australia, 35 Stirling Hwy, Crawley WA, 6009, Australia}

\author{Mark Reynolds\orcidlink{0000-0002-5415-0544}}
\affiliation{Centre for Quantum Information, Simulation and Algorithms, The University of Western Australia, 35 Stirling Hwy, Crawley WA, 6009, Australia}
\affiliation{School of Physics, Mathematics and Computing, The University of Western Australia, 35 Stirling Hwy, Crawley WA, 6009, Australia}

\author{Jingbo Wang\orcidlink{0000-0001-7544-0084}}
\affiliation{Centre for Quantum Information, Simulation and Algorithms, The University of Western Australia, 35 Stirling Hwy, Crawley WA, 6009, Australia}
\affiliation{School of Physics, Mathematics and Computing, The University of Western Australia, 35 Stirling Hwy, Crawley WA, 6009, Australia}


\newgeometry{left=2cm,right=2cm}

\begin{abstract}
\noindent

Illegal, unreported, and unregulated (IUU) fishing causes global economic losses of 10–25 billion USD annually and undermines marine sustainability and governance. 
Synthetic Aperture Radar (SAR) provides reliable maritime surveillance under all weather and lighting conditions, but classifying small maritime objects in SAR imagery remains challenging. 
We investigate quantum machine learning for this task, focusing on quantum kernel methods (QKMs) applied to real and complex SAR chips extracted from the SARFish dataset. 
We tackle two binary classification problems, the first for distinguishing vessels from non-vessels, and the second for distinguishing fishing vessels from other types of vessels. 
We compare QKMs applied to real and complex SAR chips against classical Laplacian, RBF, and linear kernels applied to real SAR chips. 
We restrict the comparison to be between just kernel based models so that the comparison is as fair and meaningful as possible.
Using noiseless numerical simulations of the quantum kernels, we find that with the real SAR chips, QKMs are capable of obtaining equal or better performance than the classical kernels in the best case.
However, the specific quantum kernel used to encode the complex SAR data overfits and performs poorly. 
This work presents the first application of QKMs to maritime classification in SAR imagery and offers insight into the potential and current limitations of quantum-enhanced learning for maritime surveillance.
\end{abstract}

\maketitle

\section{Introduction}
\label{sec:Introduction}

Image processing in the form of detecting and classifying objects is a mainstay of machine learning (ML).
The first part of this processing, namely detecting objects of interest, is a well-explored problem that has been tackled using a variety of techniques. 
For example, classical approaches such as constant false alarm rate detectors are widely used in the context of radar imagery~\cite{ElDarymli2013Target,Li2024Robust,Wen2024CFAR,Wang2024Novel,Faerch2023Comparison,Xie2022Ship}.
More recent work, however, has focused on applying deep learning approaches for detection~\cite{Schwegmann2016Deep,Zhang2020,Li2022Deep,Gupta2024Detection}, including with the YOLO framework~\cite{Redmon2016YOLO} and its variants~\cite{Yasir2024YOLO,Wang2024YOLOSAR,Selvam2025YoloSail}.

The second component of image processing, classification, has likewise been addressed using deep learning methods~\cite{Williams2023Contrastive,Wang2017Classification,Guan2023Fishing,Wang2018Ship}, in addition to other ML approaches such as classical kernel methods~\cite{AlHinai2025Confidence,Lang2017Ship} and ensemble-style combinations of classifiers~\cite{Lang2016Ship,Ji2013Ship,Yan2022Ship}.
An active open question in the scientific community is whether quantum algorithms can improve on classical ML techniques for such tasks, particularly with datasets where the objects of interest occupy only a small fraction of the total pixel area.

A pertinent example of this paradigm is the use of satellite-based imaging applied to boat detection and classification. 
Boat detection is of particular interest as illegal, unreported, and unregulated (IUU) fishing poses a significant threat to marine ecosystems and maritime governance worldwide. 
For context, the impacts of IUU fishing are far-reaching, contributing to the degradation of marine conservation efforts outlined in the United Nations Sustainable Development Goals~\cite{Okafor2019Illegal}, and causing substantial economic losses estimated at 10–25 billion USD annually~\cite{Agnew2009Estimating,Welch2022Hot}. 
IUU fishing also exacerbates political instabilities by fuelling piracy, which in turn threatens global shipping and international trade~\cite{Desai2021Measuring}.

To help combat these issues, Synthetic Aperture Radar (SAR) uses radio waves and advanced signal processing techniques to generate high-resolution images of the Earth's surface, regardless of lighting and weather conditions.
This enables persistent surveillance during both day and night under any atmospheric circumstances~\cite{Torres2019Remote}.
However, despite the high spatial resolution, the relative size of boats compared with the surrounding environment remains small, making the associated classification tasks particularly challenging.
To address this difficulty, recent work~\cite{Luckett2024Sarfish,Williams2023Contrastive} has explored the use of complex-valued SAR imagery, leveraging both amplitude and phase information for the purposes of classification.
This shift toward inherently complex data suggests a potential role for quantum algorithms, since quantum computers operate natively in complex Hilbert spaces and may be more compatible with learning from complex-valued features.

In this paper we focus on the second component of image processing, the classification of pre-detected maritime objects, and tackle the tasks using quantum kernel methods (QKMs)~\cite{Havlicek2019Supervised,Schuld2019Feature,Schuld2021Supervised}, which are kernel methods that use a quantum computer to evaluate a kernel function. 
Recent studies suggest that QKMs may offer computational advantages over classical ML methods~\cite{Nadim2025Comparative}, since they give us access to kernels which cannot be evaluated efficiently with a classical computer, such as those used in~\cite{Havlicek2019Supervised,Liu2021Rigorous,Huang2021Power,Wu2023Phase}. 
Prior works have investigated applications of quantum algorithms in the context of SAR data, including for the purposes of image formation~\cite{Giovagnoli2025Quantum,Salahat2025Quantum}, denoising~\cite{Mauro2024QSpeckle, Wang2024Quantum,Miller2024Quantum} and classification~\cite{Miller2023Quantum,Naik2024Quantum}.
This paper, however, provides the first study which investigates the efficacy of quantum kernel methods for classifying maritime objects.

We compare the learning performance metrics obtained with QKMs applied to both real and complex SAR imagery, with those obtained using classical Laplacian, radial basis function (RBF) and linear kernels applied to real SAR data, where all data is extracted from the SARFish dataset~\cite{Luckett2024Sarfish}.
In this work, we restrict our comparison to kernel-based models to ensure a fair and controlled evaluation within a consistent methodological framework. 
Kernel methods share similar inductive biases and are well-suited to the dataset size considered here. 
Including deep learning approaches would introduce additional confounding factors, such as architectural design and training heuristics, which fall outside the scope of this study.
We perform the comparison on two binary classification tasks: the first involves classifying detected maritime objects as vessels or otherwise (e.g. oil rigs), and the second involves classifying detected vessels as fishing vessels or otherwise (e.g. cargo ships).

The paper proceeds as follows. 
In Section~\ref{sec:Related Work} we discuss related works.
In Section~\ref{sec:Methods} we describe kernel methods, both classical and quantum, including the kernel-based ML algorithms that we apply in this work. In Section~\ref{sec:Experimental setup} we describe the configuration of our experiments, including the datasets, preprocessing, choice of classical and quantum kernels, and computational implementation details.
In Section~\ref{sec:Results and discussion} we provide the results of our experiments and discuss their implications. Finally, in Section~\ref{sec:Conclusion and Future Work} we conclude and provide suggestions about possible future research directions.

\section{Related Work}
\label{sec:Related Work}

In 2022 the SARFish dataset~\cite{Cao2022SARFish} was introduced and later made available to the public for use in a competition~\cite{Luckett2024Sarfish} to test boat detection and classification methods. 
The dataset builds off the xView3-SAR dataset~\cite{Paolo2022xView3} by providing coincident real-valued ground range detected (GRD) and complex-valued single look complex (SLC) products, whereas the xView3-SAR dataset provided just GRD products. 
The  dataset's purpose was to stimulate the use of complex-valued SAR imagery.

Unfortunately, the competition had no entrants and so the potential of the dataset, one of the largest datasets containing Sentinel-1 SAR imagery for maritime surveillance, is largely untapped. 
However, there have been other attempts at processing this dataset. 
In this section we will discuss prior studies that applied deep learning models to SAR datasets\footnote{For further details about other publicly available SAR datasets, including the OpenSARShip and FUSAR-Ship datasets, see Appendix E of~\cite{Paolo2022xView3} and Section 2.3 of~\cite{Luckett2024Sarfish}.} for classification, followed by those that employed classical kernel methods. 
Finally, we review previous work applying quantum algorithms to SAR imagery, including for the purposes of image formation, image denoising and classification. 

\subsection{Deep Learning with SAR data}

An earlier study by~\citet{Wang2018Ship} applied well-established deep learning models to a dataset with three classes consisting of 146 bulk carriers, 156 containers, and 144 oil tankers.
The authors reported high accuracies of 0.9766 (VGG-16), 0.9248 (VGG-19~\cite{Simonyan2015Very}), 0.9548 (Xception~\cite{Chollet2017Xception}), and 0.8947 (InceptionV3~\cite{Szegedy2016Rethinking}). 

More recently, \citet{Williams2023Contrastive} applied contrastive learning to the SARFish dataset tackling a 3-class classification problem with classes given by non-vessels, non-fishing vessels, and fishing vessels on both GRD and SLC data. 
They reported average $F_1$ scores of approximately $0.87$ and also concluded that complex-valued data provided no clear performance gain.

\citet{Guan2023Fishing} introduced a custom deep learning architecture named FishNet, which was applied to their newly constructed FishingVesselSAR dataset containing real SAR data. 
FishNet outperformed 27 baseline deep learning models and six advanced SAR-specific methods, achieving 0.8979 classification accuracy, which was between 0.0677 and 0.4759 higher than all other compared models. 
FishNet also outperformed all of the other models and methods in terms of precision, recall, and $F_1$-score.

Overall, deep learning models have demonstrated high performance on SAR-based classification tasks. 
However, many early studies relied on limited datasets, making it challenging to determine the robustness of their results. 

\subsection{Kernel methods with SAR data}

One of the first applications of kernel methods to maritime boat classification was \citet{Ji2013Ship}. 
This study used multiple handcrafted feature descriptors, whose outputs were combined using a support vector machine (SVM) with a radial basis function (RBF) kernel. 
Their method was applied to a small dataset of TerraSAR-X SAR imagery consisting of 250 samples across three classes including 150 bulk carriers, 50 oil tankers, and 50 container ships. 
This approach applied to the 3-class classification task returned accuracies between 0.81 and 0.95. 
Other early studies~\cite{Lang2016Ship, Lang2017Ship}, also typically using smaller datasets tasks, found low accuracies when dealing with more than 3 classes. However~\citet{Lang2016Ship} conducted an experiment with 150 carriers, 50 container ships, and 50 oil tankers and reported a classification accuracy of 0.9462.

A recent study by \citet{AlHinai2025Confidence} proposed and applied feature extraction techniques tailored for SAR ship classification and evaluated their performance using SVMs with both linear and RBF kernels. 
Their experiments were conducted on the OpenSARShip dataset using both GRD and SLC products for a 3-class classification task. 
They also considered the FUSAR-Ship dataset where they considered both 3-class (bulk carriers, container ships, and tankers) and 5-class (with the addition of cargo ships and fishing vessels) problems. 
For the OpenSARShip dataset, the highest accuracies achieved were 0.785 (GRD) and 0.719 (SLC). 
For the FUSAR-Ship dataset, accuracies of 0.774 (3-class) and 0.510 (5-class) were reported.

\citet{Yan2022Ship} presented a multi-stage ship classification framework that incorporated an ensemble of classifiers (which included an SVM) refined using a random forest. 
The training data included 8,000 Automatic Identification System (AIS) verified samples across four classes: cargo ships, tankers, fishing vessels, and passenger ships. 
The model was tested on non-AIS SAR data, partly constructed by the authors and partly drawn from the FUSAR-Ship dataset. In isolation, SVMs achieved accuracies of 0.7425 and 0.7575 depending on the feature set used, while the full ensemble model yielded accuracies of 0.8550 and 0.8725.

Together, these studies highlight the continued relevance and versatility of kernel-based approaches for SAR image classification. 
Despite the growing prevalence of deep learning methods, kernel methods can still perform competitively across a range of tasks and datasets. 
Moreover, their suitability for small datasets make them an attractive option for applications where data scarcity or other domain-specific constraints pose challenges. 

\subsection{Quantum algorithms with SAR}

Recent studies have turned to applying quantum algorithms to SAR data. 
Initial work has focused on the processing needed to produce SAR data. 
Two concurrent studies~\cite{Giovagnoli2025Quantum,Salahat2025Quantum} proposed quantum versions of the classical Range-Doppler Algorithm (RDA), the standard method for SAR image formation. 
Both leverage the Quantum Fourier Transform (QFT) to replace classical fast Fourier transforms (FFT), achieving a reduced runtime complexity of $\mathcal{O}(N)$ versus the classical $\mathcal{O}(N \log N)$, where $N$ is the total number of pixels in the output image. 
These results suggest potential quantum advantages in the raw processing stage of SAR pipelines. 

In the area of quantum-assisted SAR image denoising, multiple approaches have emerged. QSpeckleFilter~\cite{Mauro2024QSpeckle} introduced a QML-based speckle filtering model that targeted one of the key noise sources in SAR data. 
Similarly, Wang et al.~\cite{Wang2024Quantum} proposed a quantum morphological filtering algorithm, which performed grayscale morphological operations on all pixels simultaneously, aiming to suppress noise more efficiently than classical counterparts. 

Quantum approaches for SAR image classification tasks have also been recently explored \cite{Schmitt2018SEN12, Miller2024Quantum, Miller2023Quantum}. 
For example, Naik et al.~\cite{Naik2024Quantum} used a hybrid quantum-classical model on the 10-class MSTAR dataset, comparing a 1-layer quanvolution~\cite{Henderson2020Quanvolutional} plus CNN architecture with a purely classical 2-layer CNN. 
The quantum model achieved 0.95 accuracy, close to the 0.96 of the classical model, suggesting that shallow hybrid quantum models can remain competitive.

Overall, while quantum approaches have begun to appear in SAR research, they remain largely exploratory. 
The focus has been on accelerating classical processes (e.g., via QFT) or integrating quantum components into hybrid neural networks. 
To our knowledge, no work to date has applied quantum kernel methods to SAR imagery tasks. 
This leaves open a promising direction which we consider in this work.

\section{Methods}
\label{sec:Methods}

In this section, we begin with an overview of general kernel methods, highlighting the key structures and theorems that underpin their formulation. 
We then describe the specific kernel-based ML algorithms used in this study, and finish the section with a discussion of quantum kernel methods and the roles that quantum computers play in these algorithms.
For further details about quantum computing, we refer the reader to Chapters 1, 2 and 4 of~\cite{Nielsen2000Quantum}.

\subsection{Kernel methods}

\begin{figure*}[ht]
\centering
\begin{tikzpicture}
\draw (0.5-0.75,0.5-0.125+0.25) -- ++(2,0) -- ++(1,1) -- ++(-2,0) -- ++(-1,-1) ;
\node[rectangle, fill=blue, inner sep = 0.5mm] at (1.75,1.475) {};
\node[regular polygon, regular polygon sides=3, fill=red, inner sep = 0.25mm] at (0.85,1.3) {};
\node[regular polygon, regular polygon sides=3, fill=red, inner sep = 0.25mm] at (1.75+0.5-0.75,1.225) {};
\node[rectangle, fill=blue, inner sep = 0.5mm] at (0.55+0.5-0.75,0.915) {};
\node[regular polygon, regular polygon sides=3, fill=red, inner sep = 0.25mm] at (2.35+0.5-0.75,1.475) {};
\node[regular polygon, regular polygon sides=3, fill=red, inner sep = 0.25mm] at (0.75+0.65-0.75,1.175) {};
\node[rectangle, fill=blue, inner sep = 0.5mm] at (1.5+0.5-0.75,0.975) {};
\node[rectangle, fill=blue, inner sep = 0.5mm] at (1.25+0.5-0.75,1.525) {};
\node[regular polygon, regular polygon sides=3, fill=red, inner sep = 0.25mm] at (1.5+0.5-0.75,1.325) {};
\node[regular polygon, regular polygon sides=3, fill=red, inner sep = 0.25mm] at (2.2+0.5-0.75,1.175) {};
\node[rectangle, fill=blue, inner sep = 0.5mm] at (1.2+0.5-0.75,1.065) {};
\node[regular polygon, regular polygon sides=3, fill=red, inner sep = 0.25mm] at (0.9+0.5-0.75,0.965) {};
\node[rectangle, fill=blue, inner sep = 0.5mm] at (1.85+0.5-0.75,0.825) {};
\node at (1.75+0.5-0.75-0.75,2-0.125-1.75+0.25) {\(\mathcal{X}\equiv\R^d\)};
\draw[dotted] (5+0.125,-0.5) -- ++(1,1) -- ++(0,2) -- ++(2,0) -- ++(0,-2) -- ++(-2,0) ;
\draw (5+0.125,-0.5) -- ++ (2,0) -- ++(1,1) -- ++(0,2) -- ++(-2,0) -- ++(-1,-1) -- ++ (2,0) -- ++(1,1) -- ++(-1,-1) -- ++(0,-2) -- ++(-2,0) -- ++(0,2) ;
\node[regular polygon, regular polygon sides=3, fill=red, inner sep = 0.25mm] at (1.5+5+0.125-0.75,0.5+0.25) {};
\node[regular polygon, regular polygon sides=3, fill=red, inner sep = 0.25mm] at (1.25+5+0.125-0.75,0.4+0.25) {};
\node[regular polygon, regular polygon sides=3, fill=red, inner sep = 0.25mm] at (1.5+5+0.125-0.5,0.25+0.25) {};
\node[regular polygon, regular polygon sides=3, fill=red, inner sep = 0.25mm] at (2.2+4.25,0.35) {};
\node[regular polygon, regular polygon sides=3, fill=red, inner sep = 0.25mm] at (1.2+5,-0.25+0.25) {};
\node[regular polygon, regular polygon sides=3, fill=red, inner sep = 0.25mm] at (0.9+5-0.25,-0.05+0.25) {};
\node[regular polygon, regular polygon sides=3, fill=red, inner sep = 0.25mm] at (1.85+5,0.25) {};
\draw[thick,opacity=0.35,fill=gray] (5+0.125,0.675) -- ++(2,-0.25) -- ++(1,1) -- ++(-2,0.25) -- ++(-1,-1) ;
\draw[-stealth] (6.625,1.05) --++ (0.075*0.675,0.5*0.675);
\node at (6.75,0.85) {\fontsize{6}{7}\selectfont$\phi(x^\prime)$};
\draw[line width=0.1mm] (6.625-2*0.05,1.05+0.25*0.05) -- (6.625+2*0.05,1.05-0.25*0.05);
\draw[line width=0.1mm] (6.625-1*0.0625,1.05-1*0.0625) -- (6.625+1*0.0625,1.05+1*0.0625);
\node[rectangle, fill=blue, inner sep = 0.5mm] at (2+5+0.125,2.1) {};
\node[rectangle, fill=blue, inner sep = 0.5mm] at (2+5+0.125-0.7,2.1-0.1075) {};
\node[rectangle, fill=blue, inner sep = 0.5mm] at (1.75+5+0.125,1.85) {};
\node[rectangle, fill=blue, inner sep = 0.5mm] at (0.55+5+0.75,1.54) {};
\node[rectangle, fill=blue, inner sep = 0.5mm] at (2.35+5+0.125,1.65) {};
\node[rectangle, fill=blue, inner sep = 0.5mm] at (0.75+5.15+0.875,1.65) {};
\node at (6.75-0.75+0.125,2.875-3.75) {\(\mathcal{F}\)} ;
\draw[thick] (11-0.25+0.5+0.25,0) -- ++(0,2.5) ;
\node[rectangle, fill=blue, inner sep = 0.5mm] at (11-0.25+0.5+0.25,2.35) {};
\node[rectangle, fill=blue, inner sep = 0.5mm] at (11-0.25+0.5+0.25,2.225) {};
\node[rectangle, fill=blue, inner sep = 0.5mm] at (11-0.25+0.5+0.25,2.1) {};
\node[rectangle, fill=blue, inner sep = 0.5mm] at (11-0.25+0.5+0.25,1.89) {};
\node[rectangle, fill=blue, inner sep = 0.5mm] at (11-0.25+0.5+0.25,1.65) {};
\node[rectangle, fill=blue, inner sep = 0.5mm] at (11-0.25+0.5+0.25,1.45) {};
\draw[thick] (11.4,1.175) --++ (0.2,0);
\node at (11.7,1.175) {\scriptsize 0};
\node[regular polygon, regular polygon sides=3, fill=red, inner sep = 0.25mm] at (11-0.25+0.5+0.25,0.9) {};
\node[regular polygon, regular polygon sides=3, fill=red, inner sep = 0.25mm] at (11-0.25+0.5+0.25,0.825) {};
\node[regular polygon, regular polygon sides=3, fill=red, inner sep = 0.25mm] at (11-0.25+0.5+0.25,0.625) {};
\node[regular polygon, regular polygon sides=3, fill=red, inner sep = 0.25mm] at (11-0.25+0.5+0.25,0.5) {};
\node[regular polygon, regular polygon sides=3, fill=red, inner sep = 0.25mm] at (11-0.25+0.5+0.25,0.4) {};
\node[regular polygon, regular polygon sides=3, fill=red, inner sep = 0.25mm] at (11-0.25+0.5+0.25,0.26) {};
\node[regular polygon, regular polygon sides=3, fill=red, inner sep = 0.25mm] at (11-0.25+0.5+0.25,0.2) {};
\node at (11-0.25+0.5+0.25,2.75-3.25+0.25) {\(\R\)};
\draw[thick,-stealth ] (3.25-0.3,1.75-0.5-0.25) to[out=-30,in=-150] ++(1.75,0);
\node at (4.0725-0.3,1.5-0.5-0.5-0.25) {\(\phi(\cdot)\)};
\draw[thick,-stealth] (8.5+0.25,1.75-0.5-0.25+0.1) to[out=-30,in=-150] ++(1.75,0);
\node at (9.5+0.25,1.5-0.5-0.5-0.25+0.1) {\(\langle\phi(\cdot),\phi(x^\prime)\rangle_{\mathcal{F}}\)};
\draw[thick,-stealth ] (3-1,2.125-0.25) to[out=30,in=145] ++(9,0);
\node at (6.5,4.25-0.4-0.25) {\(\mathcal{K}(\cdot,x^\prime)\)};
\end{tikzpicture}
\caption{\footnotesize \justifying A kernel \(\mathcal{K}\), which implicitly computes an inner-product in a high-dimensional feature space \(\mathcal{F}\), can be used to simplify a binary classification problem if the associated feature map \(\phi\) arranges the inputs in \(\mathcal{F}\) in a desirable way. 
For example, above we see input data samples belonging to two different classes (shown as red triangles and blue squares) being arranged in $\mathcal{F}$ in such a way that allows a separating hyperplane to be found. 
This allows the class label associated with the points to be extracted by simply projecting along some axis in \(\mathcal{F}\) and taking the sign of the resulting value.
In the situation shown in the figure, this axis is $\phi(x^\prime)$, however in the more general case the choice of this axis is usually determined as a linear combination of the form $\sum_{i=1}^{M}\alpha_i\phi(\mathbf{x}_i)$ where the $\mathbf{x}_i$'s are the input training datapoints (see Equation~\eqref{eq:RTModelInnerProduct}).}
\label{fig:Kernel_Methods}
\end{figure*}

Kernel methods~\cite{Scholkopf2001Kernels} are a class of ML algorithms used to capture intricate patterns in moderately sized datasets. 
In principle kernel methods can be applied to datasets of arbitrary size, but the training costs scale quadratically in the number of training data points, making it difficult to scale beyond a few thousand training samples.

The key idea underlying the use of kernel methods is that of a kernel function, which implicitly computes inner products between embeddings of input data in a high (possibly infinite) dimensional feature space. 
By mapping the input data into the high-dimensional feature space, non-linear structures in the original data sometimes translate into simpler linear structures. 
This can aid the process of learning, for example, by making it easier for a ML algorithm to determine the boundary between classes in a binary classification problem (see Figure~\ref{fig:Kernel_Methods}).
Additionally, in many cases kernel methods reap the benefits of allowing an optimal solution to be found via a deterministic procedure (assuming fixed hyperparameters), which allows us to avoid complications that may arise when utilising variational ML models.

To formalise this, consider a training dataset $\mathcal{D}=\{(\mathbf{x}_i,y_i)\}_{i=1}^{M}\subset\mathcal{X}\times\{\pm1\}$ for some binary classification task. 
Here \(\mathcal{X}\equiv\F^d\) (with $\F=\C$ or $\F=\R$) is the input data domain of dimension \(d\in\mathbb{N}\), \(\mathbf{x}_i\in\mathcal{X}\) is the \(i^{\textrm{th}}\) input training data sample, \(y_i\in\{\pm1\}\) is the class label for the \(i^{\textrm{th}}\) training data sample, and \(M\in\N\) is the total number of training data samples.
A \emph{kernel} is then a symmetric function \(\mathcal{K}:\mathcal{X}\times\mathcal{X}\to\R\) such that the \emph{Gram matrix} \(K_{ij}\equiv\mathcal{K}(x_i,x_j)\) of \(\mathcal{K}\) is positive semi-definite for all choices of the set \(\{x_1,\ldots,x_m\}\subset\mathcal{X}\) and all \(m\in\N\). 
Note that here we use boldfaced $\mathbf{x}$ to denote training data samples, and $x$ to denote arbitrary elements of $\mathcal{X}$.

Any kernel \(\mathcal{K}\) can be expressed in the form
\begin{equation}
\label{eq:FeatureMapKernel}
\mathcal{K}(x,x^\prime)=\left\langle\phi(x),\phi(x^\prime)\right\rangle_{\mathcal{F}},
\end{equation}
for some function \(\phi:\mathcal{X}\to\mathcal{F}\) called a \emph{feature map}, whose codomain \(\mathcal{F}\) is a Hilbert space over \(\R\) called a \emph{feature space}.
Kernel functions can hence be viewed as implicitly calculating inner products between embeddings of inputs in a (usually high-dimensional) feature space.

Every kernel $\mathcal{K}$ also uniquely determines a Hilbert space over $\R$ known as the \emph{reproducing kernel Hilbert space} (RKHS) associated with $\mathcal{K}$.
We denote this space by $\mathcal{R}_{\mathcal{K}}$.
The RKHS $\mathcal{R}_{\mathcal{K}}$ contains functions mapping $\mathcal{X}\to\R$ and is formally defined as the completion of the real linear span of functions given by the kernel with its second argument fixed,
\begin{equation}
\label{RKHS}
\mathcal{R}_{\mathcal{K}}\equiv\overline{\textrm{span}}_{\R}\big\{\mathcal{K}(\cdot,x)|x\in\mathcal{X}\big\}.
\end{equation}
In kernel-based supervised ML, the task usually reduces to selecting an appropriate function in $\mathcal{R}_{\mathcal{K}}$ which minimises a regularised empirical risk functional on the training data.
So how can one explicitly find such a function?

In general, elements of $\mathcal{R}_{\mathcal{K}}$ do not admit representations as real linear combinations of finitely many elements in $\{\mathcal{K}(\cdot,x)|x\in\mathcal{X}\big\}$.
However, under some fairly unrestrictive conditions on the input data domain $\mathcal{X}$ and the kernel $\mathcal{K}$~\cite[Lemma 4.33]{Steinwart2008Support}, the powerful \emph{representer theorem}~\cite{Scholkopf2001Kernels,Mohri2018Foundations} applies.
In such cases, the theorem ensures that for many common learning problems, minimisers of the regularised empirical risk can be written as a finite linear combination of the form
\begin{equation}
\label{eq:RTmodel}
\mathscr{F}(\cdot)=\sum_{i=1}^{M}\alpha_i\mathcal{K}(\cdot,\mathbf{x}_i),
\end{equation}
where the coefficients \(\{\alpha_i\}_{i=1}^{M}\) are real numbers and \(\{\mathbf{x}_i\}_{i=1}^{M}\) are the input training data samples.
Thus, rather than searching over the entire high (possibly infinite) dimensional RKHS, one can instead optimise over the finite-dimensional subspace specified by the vector $\bm{\alpha}=(\alpha_1,\ldots,\alpha_M)\in\R^M$.
This finite representation enables practical computation of the solution using theoretically deterministic algorithms, as is the case with both support vector classification (SVC) and kernel ridge classification (KRC).

As a final note, by substituting~\eqref{eq:FeatureMapKernel} into~\eqref{eq:RTmodel}, we have that the minimisers of the regularised empirical risk functional can be written as
\begin{equation}
    \label{eq:RTModelInnerProduct}
    \mathscr{F}(\cdot)=\left\langle\phi(\cdot),\sum_{i=1}^{M}\alpha_i\phi(\mathbf{x}_i)\right\rangle_{\mathcal{F}}.
\end{equation}
Equation~\eqref{eq:RTModelInnerProduct} shows us that the models we obtain from kernel methods (up to a possible bias term, and taking the sign to predict class labels) are given by first embedding the argument into $\mathcal{F}$ using the feature map, and then returning the inner product of the embedded embedded argument with the vector $\sum_{i=1}^{M}\alpha_i\phi(\mathbf{x}_i)$.

\subsection{Support vector classification}
\label{sec:Support vector classification}

Support vector machines (SVMs), such as SVC, are widely regarded as some of the most successful algorithms in machine learning, especially for tackling non-linear classification problems such as those considered in this work. 
The core principle underlying SVC is based on finding a hyperplane in $\mathcal{F}$ that maximises the margin between two classes. 
This contributes to improving generalisation and robustness against noisy training data, such as the data obtained with SAR imagery.
By leveraging kernel functions, SVC can effectively capture both linear and non-linear patterns, making it a flexible and broadly applicable method across various domains.
However, despite its advantages, precisely determining the runtime complexity for SVC can be difficult.

When using a training dataset \(\mathcal{D}=\{(\mathbf{x}_i,y_i)\}_{i=1}^{M}\subset\mathcal{X}\times\{\pm1\}\), SVC involves searching for the solution to the following convex quadratic program, called the soft-margin dual optimisation problem:
\begin{equation}
\begin{aligned}
\label{SVC}
&\min_{{\bm{\alpha}}\in[0,C]^M}\frac{1}{2}\sum_{i,j=1}^{M}{\alpha}_i{\alpha}_jy_iy_jK_{ij}-\sum_{i=1}^{M}{\alpha}_i\\
&\qquad\qquad\textrm{s.t.}\quad\sum_{i=1}^{M}{\alpha}_iy_i=0.
\end{aligned}
\end{equation}
Here \(C\geq0\) is a regularisation parameter specifying the penalty associated with an incorrectly classified data point, \(K_{ij}=\mathcal{K}(\mathbf{x}_i,\mathbf{x}_j)\) is an \(M\times M\) matrix called the \emph{kernel matrix}, and the solution is the vector \({\bm{\alpha}}=(\alpha_1,\ldots,\alpha_M)\in[0,C]^M\). 
Finding a solution to Equation~\eqref{SVC} corresponds to finding a hyperplane which maximises the margin between the classes in $\mathcal{F}$, while allowing for incorrect classifications of some data points at a cost proportional to \(C\). 
This means that a solution is permitted even when the classes are not linearly separable after being embedded in \(\mathcal{F}\).  

Solving Equation~\eqref{SVC} can be achieved deterministically (up to the choice of strategy for selecting pairs of dual coefficients) using the sequential minimisation optimisation (SMO) algorithm~\cite{Platt1998Sequential, Scholkopf2001Kernels}. Classification prediction for new data, \(x\in\mathcal{X}\), can be achieved using the model 
\begin{equation}
\label{SVCModel}
f(x) = \textrm{sign}\left(\sum_{i=1}^{M}\alpha_iy_i\mathcal{K}(x,\mathbf{x}_i)+b\right),
\end{equation}
where \(b\in\mathbb{R}\) can be determined with the Karush-Kuhn-Tucker conditions~\cite{Kuhn2014Nonlinear,Karush2014Minima}. 

The SMO algorithm used to solve Equation~\eqref{SVC} has a runtime that scales somewhere between $\mathcal{O}(M)$ and $\mathcal{O}(M^2)$ with respect to the size of the dataset $M$~\cite{Platt1998Sequential}, but this does not account for specific aspects of the problem at hand. For example, the fact that we need to calculate the symmetric $M\times M$ kernel matrix incurs a runtime of $\mathcal{O}(M^2)$. Even though the SMO algorithm is guaranteed to converge in theory, large values of the regularisation parameter $C$ often incur significantly longer runtimes by leading to ill-conditioned problems that make it difficult for the SMO algorithm to converge.
This, and other factors, such as the specific solver and kernel being employed, makes precisely quantifying the runtime complexity of SVC a non-trivial task.
The lack of clear runtime complexity bounds motivates us to also consider another kernel-based ML algorithm for classification in this work, which we now discuss.

\subsection{Kernel ridge classification}
\label{sec:Kernel ridge classification}

Kernel ridge regression (KRR) is another widely successful classical machine learning method which involves seeking a function in the RKHS of a given kernel that minimises a regularised least-squares objective function.
Kernel ridge classification (KRC), more commonly known as regularised least-squares classification~\cite{Rifkin2003Regularized}, adapts KRR for binary classification by treating class labels as continuous real numbers taking on the values $\pm1$. 
Unlike many modern algorithms that rely on iterative or approximate procedures, KRC inherits from KRR the desirable property of admitting a unique closed-form solution, resulting from the convexity of its objective function. 
Once the solution has been determined, predictions for new inputs are obtained by taking the sign of the predicted continuous labels for those inputs.
Compared with SVC, KRC is conceptually simpler, requiring only the solution of a linear system, and has been empirically observed to perform comparably or better, often with lower computational costs~\cite{Fung2001Proximal}.
Additionally, the runtime complexity of KRC is much more clear cut than for SVC, further motivating the algorithm.

As with SVC, we consider a training dataset \(\mathcal{D}=\{(\mathbf{x}_i,y_i)\}_{i=1}^{M}\subset\mathcal{X}\times\{\pm1\}\).
Applying KRC to this dataset then involves finding the solution to the convex minimisation problem:
\begin{equation}
\label{KRC}
\min_{{\bm{\alpha}}\in\R^M}\sum_{i=1}^{M}\left(y_i-\sum_{j=1}^{M}\alpha_jK_{ij}\right)^2+\lambda\sum_{p,q=1}^{M}\alpha_pK_{pq}\alpha_q,
\end{equation}
where \(\lambda>0\) is a regularisation parameter specifying the penalty associated with increasing the norm of the resultant function in the RKHS. 

As mentioned, we can determine the unique vector \({\bm{\alpha}}\in\R^M\) which solves Equation~\eqref{KRC} using standard techniques from convex optimisation (see Section 11.3.2 in~\cite{Mohri2018Foundations}). 
Specifically, the vector which solves Equation~\eqref{KRC} is given by
\begin{equation}
    \label{KRCSolution}
    \bm{\alpha} = (K+\lambda \mathbb{I})^{-1}\bm{y},
\end{equation}
where $\bm{\alpha}=(\alpha_1,\ldots,\alpha_M)\in\R^M$ is the solution, $\mathbb{I}$ is an $M\times M$ identity matrix, and $\bm{y}=(y_1,\ldots,y_M)\in\{\pm1\}^{\times M}$ is the vector of labels for the training data samples.
Note that since the kernel matrix $K$ is positive semi-definite, the matrix $K+\lambda\mathbb{I}$ will always be invertible as long as $\lambda>0$. 
And the larger the value of $\lambda$ (i.e. the more regularisation), the more well-conditioned the matrix $K+\lambda\mathbb{I}$ will be, hence improving the numerical stability of the algorithm.

Once the solution $\bm{\alpha}\in\R^M$ has been determined according to Equation~\eqref{KRCSolution}, predictions are made using
\begin{equation}
    \label{KRCModel}
    g(x)=\textrm{sign}\left(\sum_{i=1}^{M}\alpha_i\mathcal{K}(x,\mathbf{x}_i)\right).
\end{equation}
Notice that determining $\bm{\alpha}$ using the right hand side of Equation~\eqref{KRCSolution} involves matrix inversion of the $M\times M$ matrix $K+\lambda\mathbb{I}$, which generally takes $\mathcal{O}(M^3)$ time. 
However we can equivalently determine $\bm{\alpha}$ by solving the system of linear equations $(K+\lambda\mathbb{I})\bm{\alpha}=\bm{y}$ which can be achieved in $\mathcal{O}(M^{2.373})$ time~\cite{LeGall2014Powers}.
So we see that, in contrast with SVC, the runtime complexity of KRC is clear.

To conclude this section, we offer some brief remarks on Kernel Ridge Classification (KRC). 
The use of a squared error loss in the first term of Equation~\eqref{KRC} is a natural choice for regression tasks, but is less well-suited to classification. 
To see this, consider an input training data sample $\mathbf{x}_j$ for some $j\in\{1,\ldots,M\}$. 
Suppose that the model prediction is large and positive, e.g. $\sum_{i=1}^{M}\alpha_i\mathcal{K}(\mathbf{x}_j,\mathbf{x}_i)=10^3$, resulting in a predicted label of $+1$ according to Equation~\eqref{KRCModel}. 
In this case, the squared loss penalty will be roughly the same regardless of whether the true label $y_j$ is +1 or -1. 
This is counter-intuitive in a classification context since the contribution to the loss function should ideally depend significantly on whether the predicted label is correct or not. 

Nonetheless, KRC has been observed to perform surprisingly well in practice. 
In~\cite{Fung2001Proximal}, the authors applied KRC to a range of benchmark datasets and found that it achieved test set accuracies comparable to those of Support Vector Machines (SVMs), while requiring significantly less computational time, up to an order of magnitude less in some cases. 
This empirical success, together with the clear runtime complexity, motivates our use of KRC in this work.

\subsection{Quantum kernel methods}
\label{sec:Quantum kernel methods}

Quantum kernel methods (QKMs)~\cite{Havlicek2019Supervised,Schuld2019Feature,Schuld2021Supervised}, are hybrid quantum-classical algorithms that use a quantum computer to compute the kernel matrix, which is then passed to a classical algorithm such as SVC or KRC.
The core idea is to prepare quantum states that depend on classical input data, effectively defining a feature map from $\mathcal{X}$ to a high-dimensional feature space, given by the Hilbert space to which the quantum states (viewed as density operators) belong.
By performing specific measurements, one can then evaluate the kernel function associated with this feature map and construct the full kernel matrix, which is later used to train a classical model. 

Formally, for all $n\in\N$ we denote by $\mathcal{H}_n$ the Hilbert space over $\R$ of $2^n\times2^n$ Hermitian matrices with the Frobenius inner-product $\langle A,B\rangle_{\mathcal{H}_n}=\textrm{tr}(A^\dagger B)$. 
An $n$-qubit \emph{quantum feature map} is then a map $\phi_Q:\mathcal{X}\to\mathcal{H}_n$, defined such that
\begin{align}
\label{QuantumFeatureMap}
\phi_Q(x)=U(x)|0\rangle\langle0|U^\dagger(x)
\end{align}
for all $x\in\mathcal{X}$. 
Here $U(x)$ is a $2^n\times2^n$ unitary matrix called the \emph{data-encoding unitary} for $x$, and $|0\rangle\in\mathbb{C}^{2^n}$ is the initial quantum state in which all qubits occupy the +1-eigenstate of the Pauli-$Z$ operator. 
Given a quantum feature map $\phi_Q$, the associated \emph{quantum kernel} $\mathcal{K}_Q:\mathcal{X}\times\mathcal{X}\to\R$, is defined by
\begin{equation}
\label{QuantumKernel}
\mathcal{K}_Q(x,x^\prime)=\left|\langle0|U^\dagger(x^\prime)U(x)|0\rangle\right|^2
\end{equation}
for all $x,x^\prime\in\mathcal{X}$. 
Note that, as in Equation~\eqref{eq:FeatureMapKernel},  $\mathcal{K}_Q(x,x^\prime)$ is just the inner-product $\langle\phi_Q(x),\phi_Q(x^\prime)\rangle_{\mathcal{H}_n}$.

If the physical implementation of $U(x)$ for all $x\in\mathcal{X}$ on a quantum computer has time complexity $\mathcal{O}(\textrm{poly}(n))$, then a quantum computer can efficiently evaluate $\mathcal{K}_Q$.
In particular, from Equation~\eqref{QuantumKernel} and the Born rule we see that $\mathcal{K}_Q(x,x^\prime)$ is just the expectation value of the observable $|0\rangle\langle0|$ measured from the pure state $U^\dagger(x^\prime)U(x)|0\rangle$. 
This provides us with one physical procedure for evaluating the kernel $\mathcal{K}_Q$ on quantum hardware\footnote{There are other methods, such as a SWAP test \cite[Section III.B]{Paine2023Quantum}.}.

Training involves evaluating all the entries of $K_{ij}=\mathcal{K}_Q(\mathbf{x}_i,\mathbf{x}_j)$, where $\{\mathbf{x}_i\}_{i=1}^{M}$ are the training data samples, and passing the matrix to a classical algorithm.
Finally, with reference to Equations~\eqref{SVCModel} and \eqref{KRCModel}, making further predictions using the trained model requires another $\mathcal{O}(M)$ evaluations of the quantum kernel.
Overall, this incurs a runtime complexity of $\mathcal{O}(M(M+N)\,\textrm{poly}(n))$ on quantum hardware, where $N$ denotes the number of testing datapoints.
However, this doesn't account for the number of measurements required to resolve quantum kernel values, which we briefly discuss in the last two paragraphs of this section.

Note that many descriptions of quantum kernel methods (QKMs), such as~\cite{Schnabel2025Scrutiny}, incorporate variational parameters. 
In contrast, we consider fixed feature maps with no variational parameters, only hyperparameters such as $C$ and $\lambda$ (and the quantum kernel bandwidth discussed in Section~\ref{sec:Kernels}) tuned via cross-validation. 
This avoids issues such as the barren plateau phenomenon associated with variational quantum models~\cite{Larocca2025Barren}, which can hinder their experimental viability. 
As a result, all training is performed classically, and the QKMs inherit the standard properties of kernel methods, including the ability to obtain deterministic solutions in the corresponding reproducing kernel Hilbert space when using SVC or KRC.

While QKMs share similarities and benefits with general kernel methods, they also present unique challenges. 
For example, they are susceptible to quantum hardware noise~\cite{Heyraud2022NoisyQKMs,Wang2021Towards,Hubregtsen2022Training}, and in some cases experience a phenomena known as exponential concentration, which imposes a required number of measurements that scales exponentially with the number of qubits~\cite{Thanasilp2024Exponential}. 
Although a detailed investigation of these challenges lies beyond the scope of this work, they are important to acknowledge as they directly impact the practical feasibility of QKMs. 

In particular, recent work has begun to characterise the measurement sample complexity of quantum kernel methods, which determines an important component of their computational cost. 
For instance, Gentinetta et al.~\cite{Gentinetta2024Complexity} show that, under certain assumptions, quantum SVMs can be trained to accuracy $\tilde{\epsilon}$ using $\mathcal{O}(M^{4.67}/\tilde{\epsilon}^2)$ measurements for a dataset of size $M$, although this analysis does not account for exponential concentration effects.
Related work by Miroszewski et al.~\cite{Miroszewski2024QKMShots} incorporates both concentration and the distribution of kernel values into estimates of the measurement cost required to evaluate quantum kernels, highlighting the importance of these effects in realistic settings. 
Taken together, these results suggest that while QKMs may offer computational advantages, their practical utility depends critically on the interplay between measurement overhead and the statistical properties of the kernel.

\section{Numerical Experimental Setup}
\label{sec:Experimental setup}

Here we discuss our training methodology. A schematic diagram illustrating the overall workflow can be found in Figure~\ref{fig:Workflow_schematic}. 
We start by describing how the SAR data is preprocessed, then discuss our workflow and the specific kernels, both classical and quantum, which we trial and numerically simulate (without noise). 
We finish with a brief description of our specific computational implementation.



\subsection{Datasets}
\label{sec:Datasets}

In this work, we make use of datasets containing chip images of maritime objects extracted from the SARFish dataset~\cite{Cao2022SARFish,Luckett2024Sarfish}. 
As discussed in Section~\ref{sec:Related Work}, the SARFish dataset is a free and open-source SAR dataset containing coincident real-valued GRD and complex-valued SLC products, both of which are derived from Sentinel-1 imagery.
The dataset consists of two subsets split based on the polarisation of the emitted microwave pulse from the satellite: so-called VV and VH. In~\cite{Luckett2024Sarfish}, the authors state that the appearance of ships is usually more prominent in the VH polarisation.
For this reason, we conduct our experiments using the VH products.

The GRD data product contains unsigned \texttt{int16} pixel values describing the real intensity of the reflected signal, with each pixel representing a physical area of 10m $\times$ 10m. 
SLC products contain complex \texttt{int16} pixel values describing amplitude and phase information of the reflected signal, with each pixel representing a physical area of 2.3m $\times$ 14.1m. 
The dataset also contains a list of detected objects for each image product. 

Each detected object has several labels, the key ones being: whether it belongs to a GRD or SLC image; the location (as a row and column value) in the image, and two relevant classification labels named \texttt{is\_vessel}, \texttt{is\_fishing}, along with the confidence of either label. 
The \texttt{is\_vessel} and \texttt{is\_fishing} labels are binary labels describing whether the detected object is a marine vessel, and whether a marine vessel is a fishing vessel (respectively). 
The \texttt{confidence} label takes on three values, HIGH, MEDIUM, or LOW, and describes the level of confidence in the \texttt{is\_vessel} and \texttt{is\_fishing} labels for that object.

Using the GRD and SLC products together with the \texttt{is\_vessel}, \texttt{is\_fishing} and \texttt{confidence} labels, we extracted 6 datasets, one for each choice of \texttt{is\_vessel} or \texttt{is\_fishing} with either $16\times16$ GRD chips, $16\times16$ SLC chips, or $70\times12$ SLC chips\footnote{Note that the $70\times12$ SLC chips corresponds to a similar surface area as the $16\times16$ GRD chips}. Some examples of the $16\times16$ GRD chips can be seen in Figure~\ref{fig:Workflow_schematic}(a), together with their associated \texttt{is\_vessel} labels, from which one can see the difficulty in visually distinguishing between the classes.

To construct the datasets, we first discard all detected objects without a \texttt{HIGH} confidence label to reduce label noise and ensure that the extracted GRD and SLC chips have reliable ground-truth annotations. 
This confidence level does not imply better image quality, it simply indicates that the labels are expected to be accurate, often supported by AIS-verified information.
Using the remaining detected objects, we then randomly sample balanced datasets containing 1250 data samples each, resulting in 625 samples per class. 
This is so that we can reserve 80\% of the data for training and still maintain 1000 training data samples, a suitable amount for applying kernel methods.

\subsection{Preprocessing}
\label{sec:Preprocessing}

Once each of the 6 datasets described in the previous subsection was extracted from the SARFish dataset, we performed three main preprocessing steps.
A visual depiction of each of the preprocessing steps can be seen in Figure~\ref{fig:Workflow_schematic}(a).

We start by applying the function $h:\C\to\C$, defined such that
\begin{equation}
\label{eq:h-preprocessing}
    h(z)=\ln(1+|z|)e^{i\arg(z)},
\end{equation}
for all $z\in\C$, to all pixel values in all chip images.
Applying $h$ to non-negative real inputs, such as the unsigned \texttt{int16} values present in the GRD chips, gives the same output as the \texttt{numpy}~\cite{Harris2020Numpy} function \texttt{log1p}. 
However $h$ also accepts complex arguments, transforming the arguments in such a way that preserves phase information and only alters their complex modulus.
We choose to perform this preprocessing step so that pixel values with (complex or real) moduli many orders of magnitude larger than others will not dominate the variance of the pixel values.

Next, we split the sampled data into training and testing data.
Specifically, we employ stratified sampling and use 80\% of the total data for training (1000 samples with 500 from each class) and the remaining 20\% for testing (250 samples with 125 from each class).

In the final preprocessing step, we flatten each chip into a vector and apply principal component analysis (PCA) on each of the training datasets to derive an orthogonal linear transformation. 
The choice of PCA for dimensionality reduction is natural in this context and has been applied in prior studies that apply quantum kernel methods to classical datasets~\cite{Canatar2023Bandwidth,Shaydulin2022Importance,Schnabel2025Scrutiny}.
The transformation derived from the training data is then applied to the associated testing datasets.

In our experiments, we trial every number of principal components from 1 up to 12, resulting in the chip images being represented by vectors of length 1 to 12, with real or complex entries depending on whether we started with GRD or SLC chips (respectively).
Our decision not to extend beyond 12 principal components was largely a result of computational restrictions. 
But also of the fact that this number of components resulted in a large portion of the variance in the data being preserved, the remainder of which we expect to be largely attributed to noisy fluctuations typical of SAR imagery.

In Appendix~\ref{sec:Principal component analysis considerations} we provide a plot of the cumulative explained variance ratios as a function of the number of principal components for the $16\times16$ GRD chips used for the \texttt{is\_vessel} and \texttt{is\_fishing} classification tasks, from which we can see that using 12 principal components preserves roughly 85\% of the total variance in these datasets.
We did not produce similar plots for the SLC datasets.
These final vectors are then used as the input training and testing data samples.

\subsection{Machine learning workflow}
\label{sec:Machine learning workflow}

\begin{figure*}
\centering
\begin{tikzpicture}

\node at (8.5,7-0.125) {\large\textbf{(a) Dataset preprocessing}};
\draw[thick, fill=blue!10, rounded corners] (0,-0.125) --++ (17,0) --++ (0,6.625) --++ (-17,0) -- cycle;

\node at (2.125,6.125) {\footnotesize Raw chip-image dataset};
\draw[thick, dotted, fill=black!1] (0.375,0.1) --++ (3.5,0) --++ (0,5.75) -| cycle;

\node at (1.25, 5.75-0.25) {\footnotesize Non-vessels};
\node at (1.25,5-0.25) {\includegraphics[width=0.95cm,trim={0.75cm 0.375cm 0.75cm 1cm},clip]{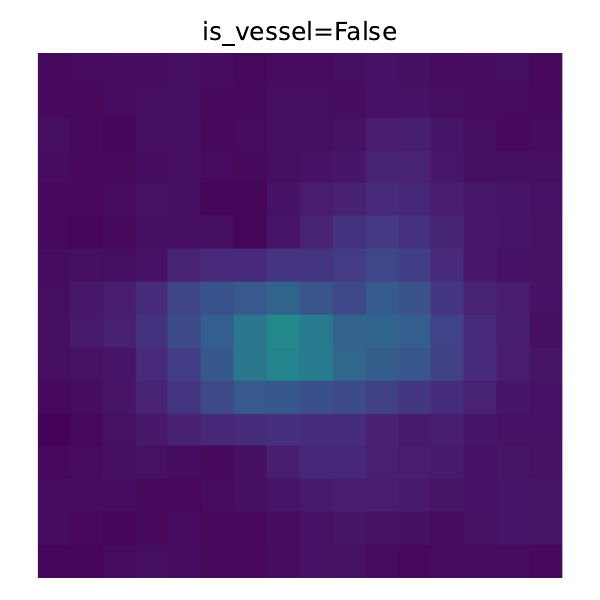}};
\node at (1.25,4-0.25) {\includegraphics[width=0.95cm,trim={0.75cm 0.375cm 0.75cm 1cm},clip]{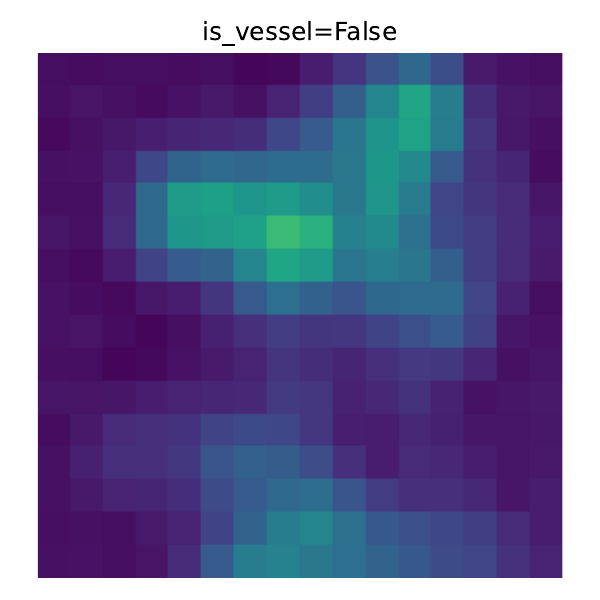}};
\node at (1.25,3-0.25) {\includegraphics[width=0.95cm,trim={0.75cm 0.375cm 0.75cm 1cm},clip]{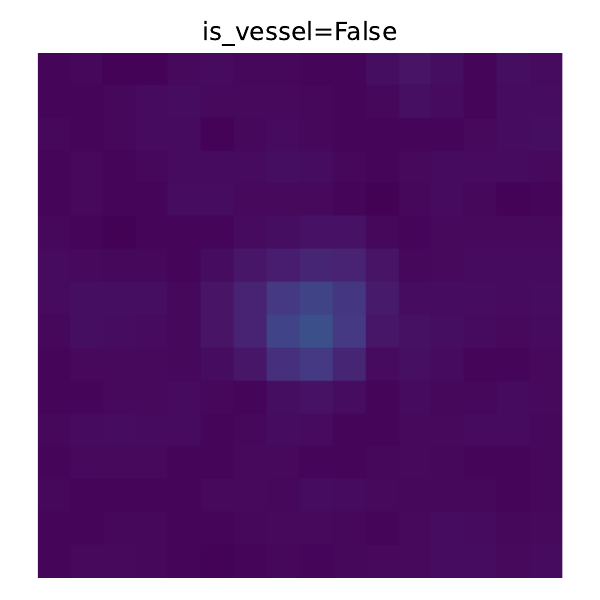}};
\node at (1.25,2-0.25) {\includegraphics[width=0.95cm,trim={0.75cm 0.375cm 0.75cm 1cm},clip]{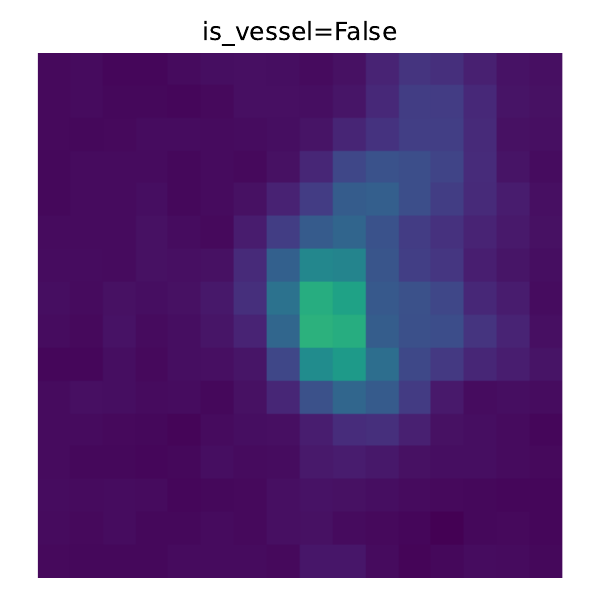}};
\node at (1.25,1-0.25) {\includegraphics[width=0.95cm,trim={0.75cm 0.375cm 0.75cm 1cm},clip]{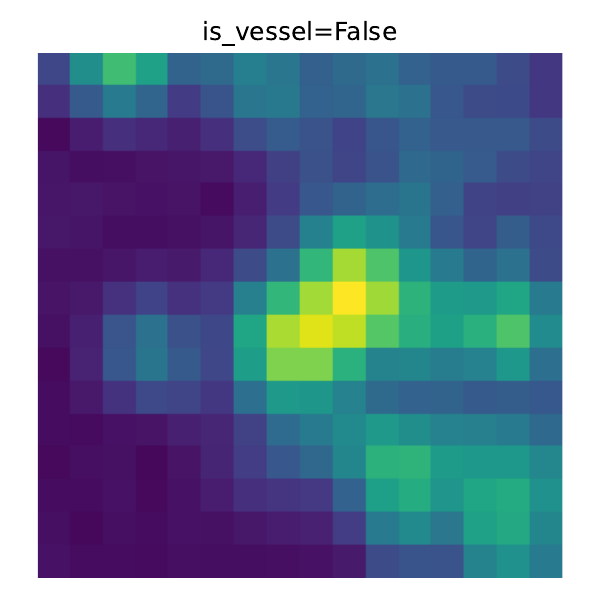}};

\node at (3,5.75-0.25) {\footnotesize Vessels};
\node at (3,5-0.25) {\includegraphics[width=0.95cm,trim={0.75cm 0.375cm 0.75cm 1cm},clip]{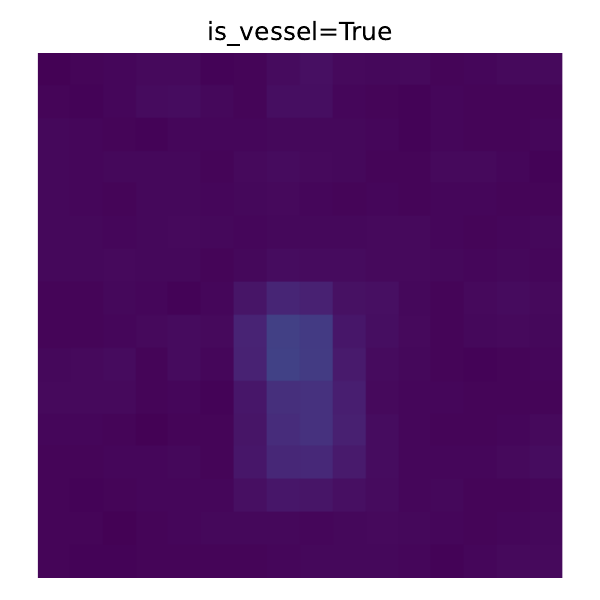}};
\node at (3,4-0.25) {\includegraphics[width=0.95cm,trim={0.75cm 0.375cm 0.75cm 1cm},clip]{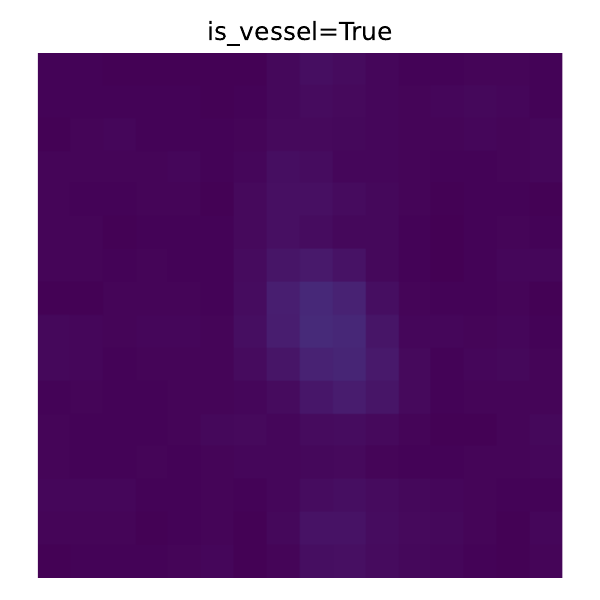}};
\node at (3,3-0.25) {\includegraphics[width=0.95cm,trim={0.75cm 0.375cm 0.75cm 1cm},clip]{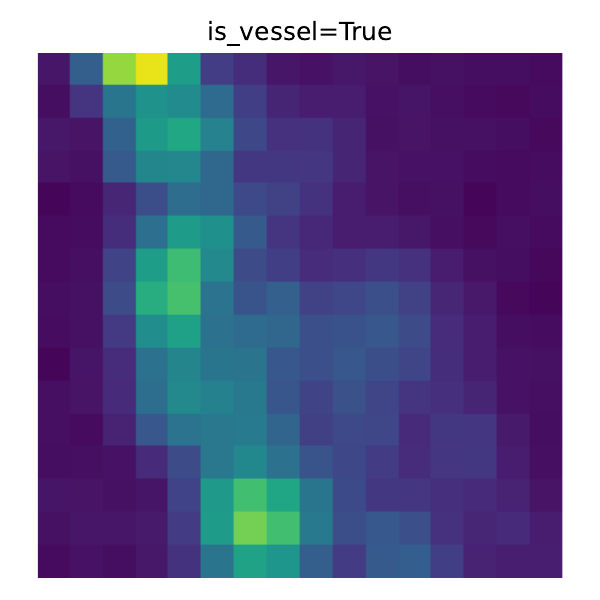}};
\node at (3,2-0.25) {\includegraphics[width=0.95cm,trim={0.75cm 0.375cm 0.75cm 1cm},clip]{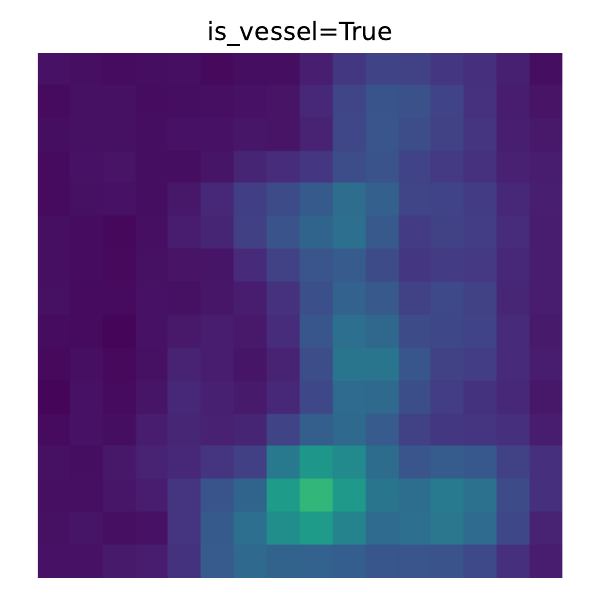}};
\node at (3,1-0.25) {\includegraphics[width=0.95cm,trim={0.75cm 0.375cm 0.75cm 1cm},clip]{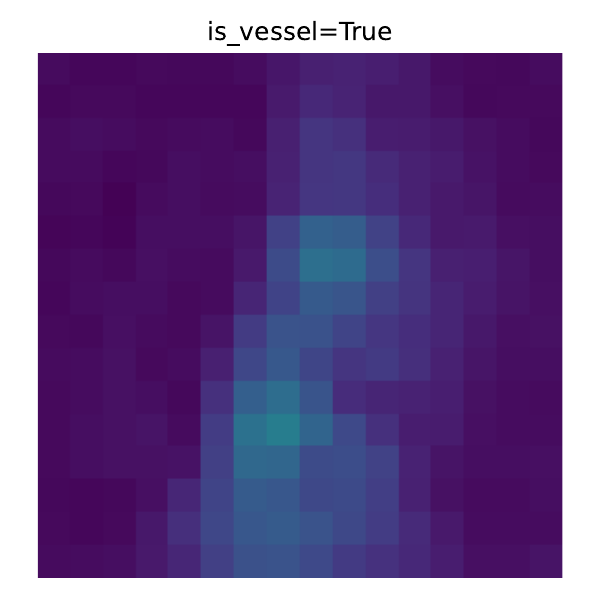}};

\node[rotate=90] at (4.2,3.05) {\tiny Apply $h$ from\,\, Equation~\eqref{eq:h-preprocessing}}; 
\draw[thick,line width=0.25mm,black,-stealth] (4,3) --++ (0.5,0);

\node at (2.125+4.25,6.125) {\footnotesize Processed chip-image dataset};
\draw[thick, dotted, fill=black!1] (4.625,0.1) --++ (3.5,0) --++ (0,5.75) -| cycle;

\node at (1.25+4.25, 5.75-0.25) {\footnotesize Non-vessels};
\node at (1.25+4.25,5-0.25) {\includegraphics[width=0.95cm,trim={0.75cm 0.375cm 0.75cm 1cm},clip]{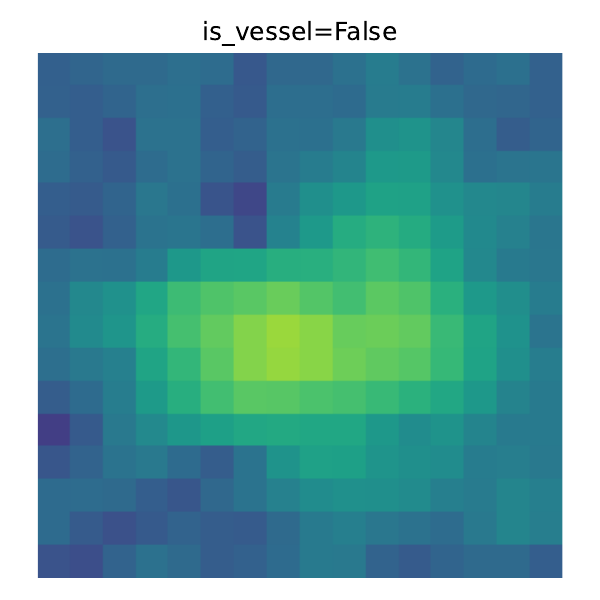}};
\node at (1.25+4.25,4-0.25) {\includegraphics[width=0.95cm,trim={0.75cm 0.375cm 0.75cm 1cm},clip]{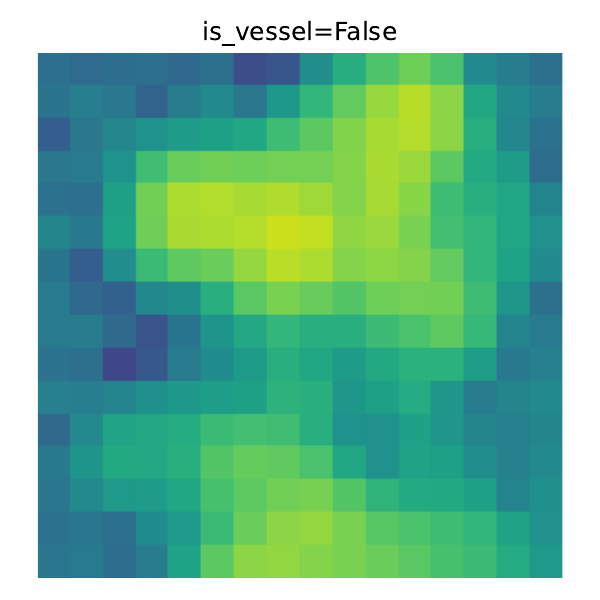}};
\node at (1.25+4.25,3-0.25) {\includegraphics[width=0.95cm,trim={0.75cm 0.375cm 0.75cm 1cm},clip]{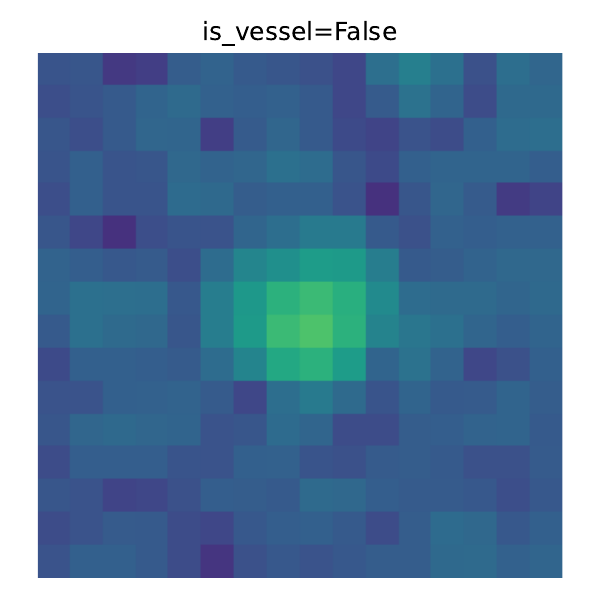}};
\node at (1.25+4.25,2-0.25) {\includegraphics[width=0.95cm,trim={0.75cm 0.375cm 0.75cm 1cm},clip]{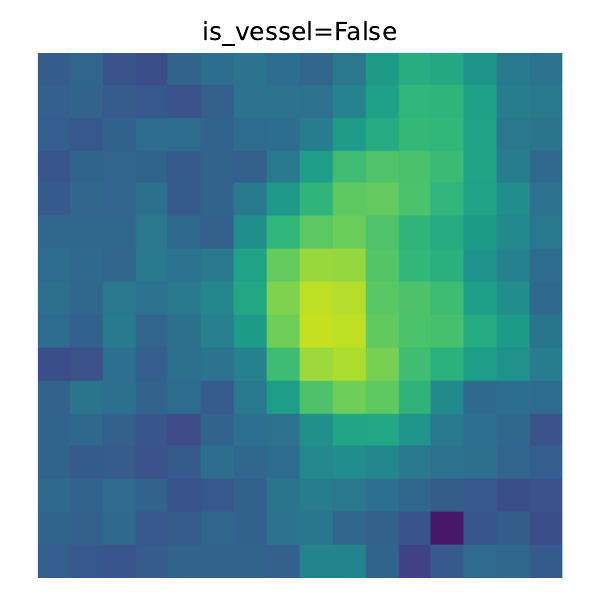}};
\node at (1.25+4.25,1-0.25) {\includegraphics[width=0.95cm,trim={0.75cm 0.375cm 0.75cm 1cm},clip]{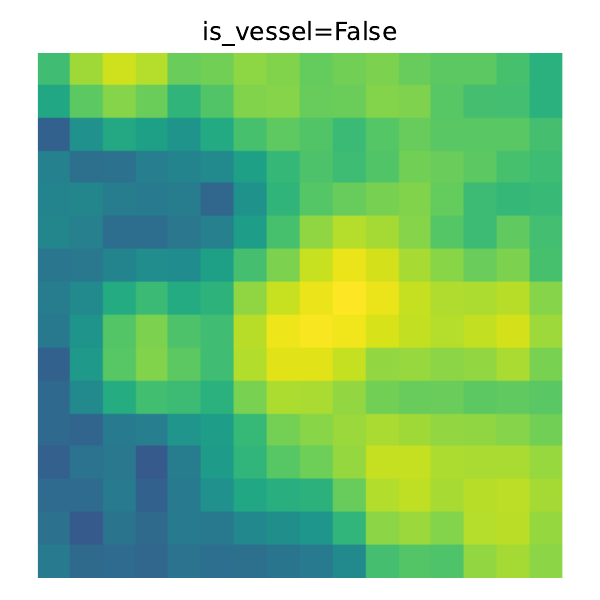}};

\node at (3+4.25,5.75-0.25) {\footnotesize Vessels};
\node at (3+4.25,5-0.25) {\includegraphics[width=0.95cm,trim={0.75cm 0.375cm 0.75cm 1cm},clip]{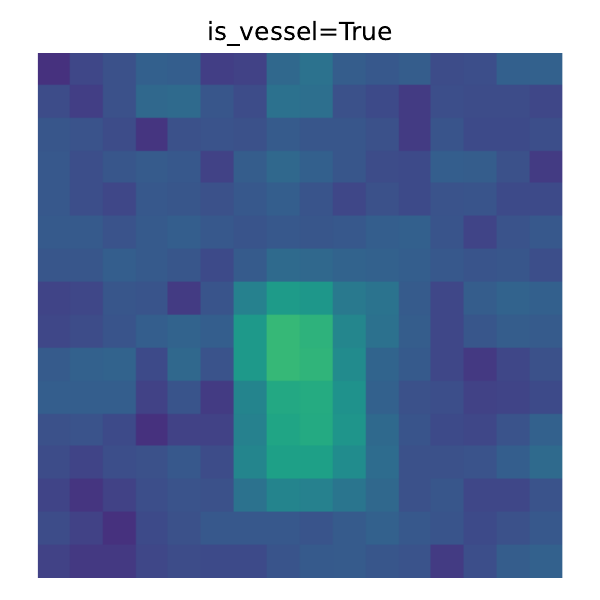}};
\node at (3+4.25,4-0.25) {\includegraphics[width=0.95cm,trim={0.75cm 0.375cm 0.75cm 1cm},clip]{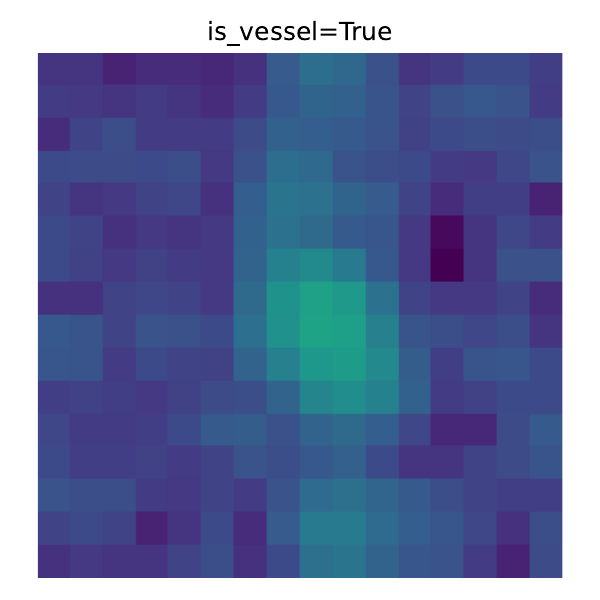}};
\node at (3+4.25,3-0.25) {\includegraphics[width=0.95cm,trim={0.75cm 0.375cm 0.75cm 1cm},clip]{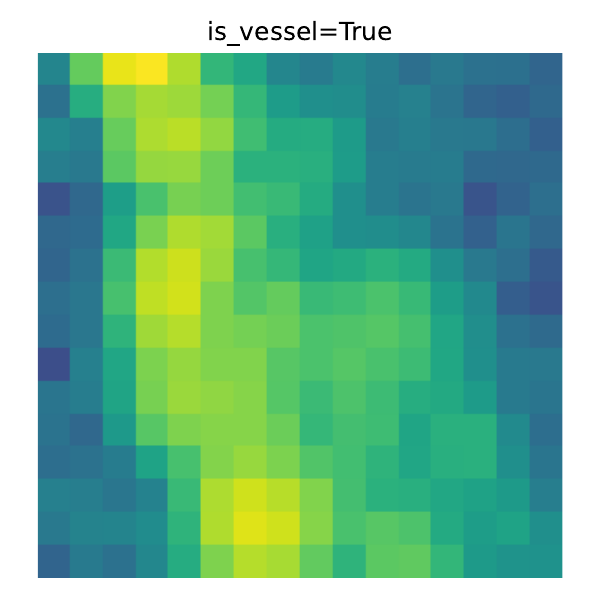}};
\node at (3+4.25,2-0.25) {\includegraphics[width=0.95cm,trim={0.75cm 0.375cm 0.75cm 1cm},clip]{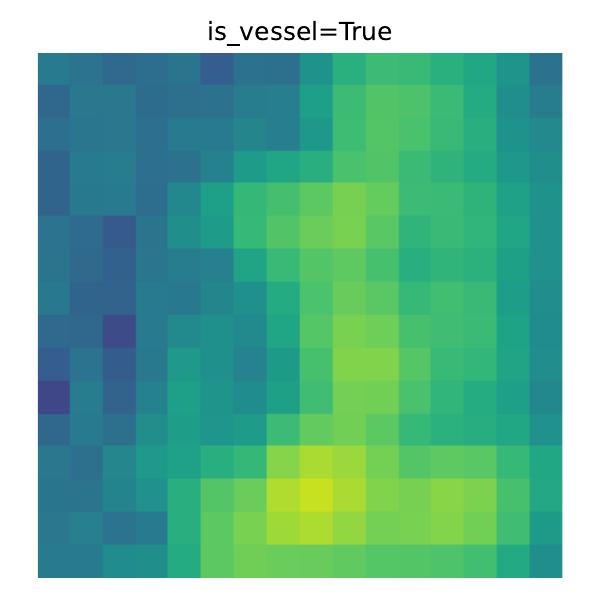}};
\node at (3+4.25,1-0.25) {\includegraphics[width=0.95cm,trim={0.75cm 0.375cm 0.75cm 1cm},clip]{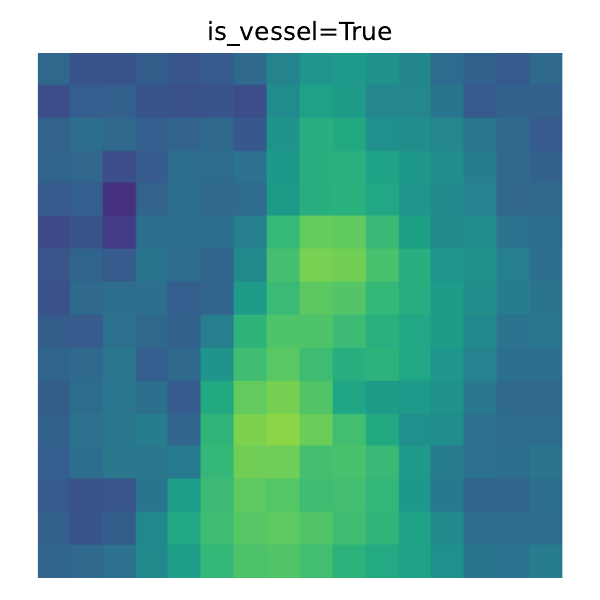}};

\node[rotate=90] at (4.2+4.3,3.15) {\tiny $\substack{\textrm{\,\,\,\,Apply stratified \,\,\,sampling to split the}\\\textrm{data into training \,\,\,\,and testing datasets}}$}; 
\draw[thick,line width=0.25mm,black,-stealth] (4+4.25,3) --++ (0.5,0);

\node at (2.125+4.25+4.25,6.25) {\footnotesize Training dataset (80\%)};
\node at (2.125+4.25+4.25,1.45) {\footnotesize Testing dataset (20\%)};
\draw[thick, dotted, fill=black!1] (4.625+4.25,1.65+0.125) --++ (3.5,0) --++ (0,4.25) -| cycle;
\draw[thick, dotted, fill=black!1] (4.625+4.25,0) --++ (3.5,0) --++ (0,1.2) -| cycle;

\node at (1.25+4.25+4.25,5+0.375) {\includegraphics[width=0.95cm,trim={0.75cm 0.375cm 0.75cm 1cm},clip]{GRD_isVessel_False_0_log1p.pdf}};
\node at (1.25+4.25+4.25,4+0.375) {\includegraphics[width=0.95cm,trim={0.75cm 0.375cm 0.75cm 1cm},clip]{GRD_isVessel_False_3_log1p.pdf}};
\node at (1.25+4.25+4.25,3+0.375) {\includegraphics[width=0.95cm,trim={0.75cm 0.375cm 0.75cm 1cm},clip]{GRD_isVessel_False_5_log1p.pdf}};
\node at (1.25+4.25+4.25,2+0.375) {\includegraphics[width=0.95cm,trim={0.75cm 0.375cm 0.75cm 1cm},clip]{GRD_isVessel_False_38_log1p.pdf}};
\node at (1.25+4.25+4.25,1-0.375) {\includegraphics[width=0.95cm,trim={0.75cm 0.375cm 0.75cm 1cm},clip]{GRD_isVessel_False_73_log1p.pdf}};

\node at (3+4.25+4.25,5+0.375) {\includegraphics[width=0.95cm,trim={0.75cm 0.375cm 0.75cm 1cm},clip]{GRD_isVessel_True_3802_log1p.pdf}};
\node at (3+4.25+4.25,4+0.375) {\includegraphics[width=0.95cm,trim={0.75cm 0.375cm 0.75cm 1cm},clip]{GRD_isVessel_True_3807_log1p.pdf}};
\node at (3+4.25+4.25,3+0.375) {\includegraphics[width=0.95cm,trim={0.75cm 0.375cm 0.75cm 1cm},clip]{GRD_isVessel_True_3813_log1p.pdf}};
\node at (3+4.25+4.25,2+0.375) {\includegraphics[width=0.95cm,trim={0.75cm 0.375cm 0.75cm 1cm},clip]{GRD_isVessel_True_3870_log1p.pdf}};
\node at (3+4.25+4.25,1-0.375) {\includegraphics[width=0.95cm,trim={0.75cm 0.375cm 0.75cm 1cm},clip]{GRD_isVessel_True_3874_log1p.pdf}};

\node[rotate=90] at (4.2+4.25+4.25,4-0.175) {\tiny 
$\substack{\textrm{\,\,Reduce \, dimensionality}\\\textrm{with principal \, component analysis}}$}; 
\draw[thick,line width=0.25mm,black,-stealth] (4+4.25+4.25,3+0.5) --++ (0.5,0);

\draw[thick, dotted, fill=black!1] (4.625+4.25+4.25,1.65+0.125) --++ (3.5,0) --++ (0,4.25) -| cycle;
\draw[thick, dotted, fill=black!1] (4.625+4.25+4.25,0) --++ (3.5,0) --++ (0,1.2) -| cycle;

\node at (1.25+4.25+4.25+4.25,5+0.375) {$\left[\begin{smallmatrix}\circ\\\circ\\\circ\\\circ\\\end{smallmatrix}\right]$};
\node at (1.25+4.25+4.25+4.25,4+0.375) {$\left[\begin{smallmatrix}\circ\\\circ\\\circ\\\circ\\\end{smallmatrix}\right]$};
\node at (1.25+4.25+4.25+4.25,3+0.375) {$\left[\begin{smallmatrix}\circ\\\circ\\\circ\\\circ\\\end{smallmatrix}\right]$};
\node at (1.25+4.25+4.25+4.25,2+0.375) {$\left[\begin{smallmatrix}\circ\\\circ\\\circ\\\circ\\\end{smallmatrix}\right]$};
\node at (1.25+4.25+4.25+4.25,1-0.375) {$\left[\begin{smallmatrix}\circ\\\circ\\\circ\\\circ\\\end{smallmatrix}\right]$};

\node at (3+4.25+4.25+4.25,5+0.375) {$\left[\begin{smallmatrix}\circ\\\circ\\\circ\\\circ\\\end{smallmatrix}\right]$};
\node at (3+4.25+4.25+4.25,4+0.375) {$\left[\begin{smallmatrix}\circ\\\circ\\\circ\\\circ\\\end{smallmatrix}\right]$};
\node at (3+4.25+4.25+4.25,3+0.375) {$\left[\begin{smallmatrix}\circ\\\circ\\\circ\\\circ\\\end{smallmatrix}\right]$};
\node at (3+4.25+4.25+4.25,2+0.375) {$\left[\begin{smallmatrix}\circ\\\circ\\\circ\\\circ\\\end{smallmatrix}\right]$};
\node at (3+4.25+4.25+4.25,1-0.375) {$\left[\begin{smallmatrix}\circ\\\circ\\\circ\\\circ\\\end{smallmatrix}\right]$};

\draw[thick,line width=0.25mm, black, -stealth] (14.75,1.675) --++(0,-0.375);
\node at (14.75+0.07, 1.475) {\tiny $\substack{\textrm{Apply derived\,\, PCA transform}\\\textrm{\,\,\,\,\,\,\,\,\,\,\,to the\,\,\, testing data}}$};
\draw[thick,line width=0.25mm,black,-stealth] (4+4.25+4.25,0.625) --++ (0.5,0);


\node at (4,-1.125) {\large\textbf{(b) Training}};
\draw[thick, fill=red!10, rounded corners] (0,-6) --++ (8.25,0) --++ (0,4.5) --++ (-8.25,0) -- cycle;

\draw[thick, rounded corners, black, dashed, line width=0.25mm, -stealth] (16.75,3.75) --++ (0.5,0) --++ (0,-4.375) --++ (-17.5,0) --++ (0,-2.25+0.125) --++ (0.5,0) ;

\node at (2,-1.75) {\footnotesize Data-encoding unitary};
\node at (2,-3) {\includegraphics[width=3.5cm,trim={1cm 0cm 0cm 0cm},clip]{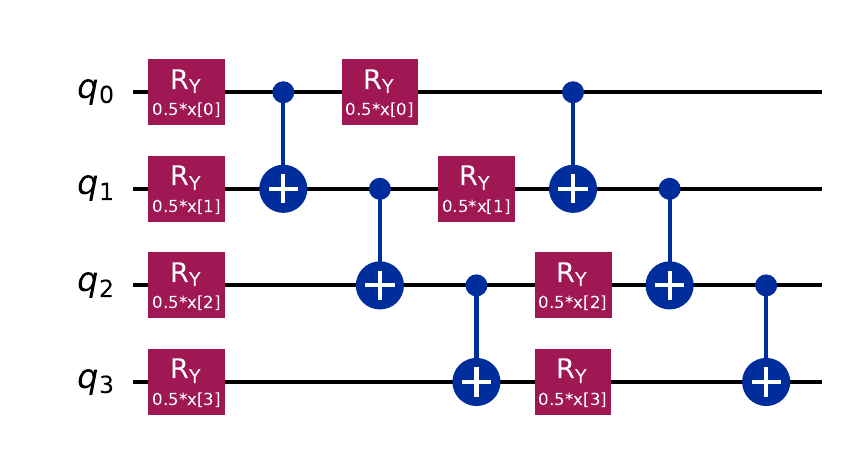}};

\node[rotate=90] at (4,-2.95) {\tiny $\substack{\textrm{Evaluate\, quantum}\\\textrm{\,\,\,\,kernel \, function}}$};
\draw[thick, line width=0.25mm, black, -stealth] (4-0.25,-3) --++ (0.5,0);

\node at (6.25,-2.125+0.335) {\footnotesize Quantum kernel matrix};
\node at (6.25,-2.125) {\footnotesize for training};
\node at (6.25,-3) {\tiny$\begin{bmatrix}\mathcal{K}_Q(\mathbf{x}_1,\mathbf{x}_1)&\dots&\mathcal{K}_Q(\mathbf{x}_1,\mathbf{x}_M)\\\vdots&\ddots&\vdots\\\mathcal{K}_Q(\mathbf{x}_M,\mathbf{x}_1)&\dots&\mathcal{K}_Q(\mathbf{x}_M,\mathbf{x}_M)\\\end{bmatrix}$};

\draw[thick, dotted, fill=black!1] (0.8,-4.75) --++ (2.5,0) --++ (0,-1) -| cycle;
\node at (2,-5.25) { $\substack{\textrm{SVC}\\\textrm{or}\\\textrm{KRC}}$};
\node at (2.125,-4.5) {\footnotesize Kernel-based algorithm};

\draw[thick, line width=0.25mm, black, -stealth] (6.175,-3.575) --++ (0,-0.675) --++ (-5.875,0) --++(0,-1) --++ (0.375,0);

\draw[thick, dotted, fill=black!1] (4.8,-4.75) --++ (2.5,0) --++ (0,-1) -| cycle;
\node at (2+4,-5.25) {$\substack{f(x)\textrm{ from Eq.~\eqref{SVCModel}}\\\textrm{or}\\g(x)\textrm{ from Eq.~\eqref{KRCModel}}}$};;
\node at (2.125+4,-4.5) {\footnotesize Trained model};

\draw[thick, -stealth, line width=0.25mm] (3.45,-5.25) --++ (1.25,0);
\node at (4,-5.235) {\tiny $\substack{\textrm{Tune hyper-}\\\textrm{parameters}\\\\\textrm{via cross-}\\\textrm{validation}}$};

\node at (13,-1.125) {\large\textbf{(c) Testing}};
\draw[thick, fill=yellow!10, rounded corners] (9-0.25,-6) --++ (8.25,0) --++ (0,4.5) --++ (-8.25,0) -- cycle;

\draw[thick, dashed, line width=0.25mm, -stealth, rounded corners] (14.75,-0.05) --++ (0,-0.35) --++ (-6.375,0) --++ (0,-2.5) --++ (0.625,0);

\draw[thick, rounded corners, black, dashed, line width=0.25mm, -stealth] (9-0.25,-0.65) --++ (-0.125,0) --++ (0,-2) --++ (0.35,0);

\node at (2+8.75,-1.75) {\footnotesize Data-encoding unitary};
\node at (2+8.75,-3) {\includegraphics[width=3.5cm,trim={1cm 0cm 0cm 0cm},clip]{Ry_1DStaircaseCx.pdf}};

\node[rotate=90] at (4+8.75,-2.95) {\tiny $\substack{\textrm{Evaluate\, quantum}\\\textrm{\,\,\,\,kernel \, function}}$};
\draw[thick, line width=0.25mm, black, -stealth] (4-0.25+8.75,-3) --++ (0.5,0);

\node at (6.25+8.75,-2.125+0.335) {\footnotesize Quantum kernel matrix};
\node at (6.25+8.75,-2.125) {\footnotesize for testing};
\node at (6.25+8.75,-3) {\tiny$\begin{bmatrix}\mathcal{K}_Q(x_1,\mathbf{x}_1)&\dots&\mathcal{K}_Q(x_1,\mathbf{x}_M)\\\vdots&\ddots&\vdots\\\mathcal{K}_Q(x_N,\mathbf{x}_1)&\dots&\mathcal{K}_Q(x_N,\mathbf{x}_M)\\\end{bmatrix}$};

\draw[thick, dashed, line width=0.25mm, -stealth] (7.5,-5.25) --++ (1.875,0) ;

\draw[thick, line width=0.25mm, black, -stealth] (6.175+8.75,-3.575) --++ (0,-0.675) --++ (-5.875,0) --++(0,-0.75) --++ (0.365,0);

\draw[thick, dotted, fill=black!1] (0.8+8.75,-4.75+0.125) --++ (2.5,0) --++ (0,-1) -| cycle;
\node at (2+8.75,-5.25+0.125) {\footnotesize$\substack{\textrm{Predict labels for}\\\textrm{ the testing data}\\\textrm{using the trained}\\\textrm{model}}$};

\draw[thick, line width=0.25mm, black, -stealth] (6.175,-3.575) --++ (0,-0.675) --++ (-5.875,0) --++(0,-1) --++ (0.375,0);

\draw[thick, dotted, fill=black!1] (4.8+8.75,-4.75+0.11) --++ (2.5,0) --++ (0,-1) -| cycle;
\node at (2+4+8.75,-5.25+0.125) {$\substack{\textrm{ Calculate learning}\\\textrm{performance metrics}\\\textrm{from the true and}\\\textrm{predicted test labels}}$};

\draw[thick, -stealth, line width=0.25mm] (3.45+8.75,-5.125) --++ (1.2,0);

\end{tikzpicture}
\caption{\footnotesize \justifying
Schematic diagram of our kernel-based machine-learning workflow with four qubits.
\textbf{(a)} Raw chip-image data is first preprocessed by applying the function $h$ from Equation~\eqref{eq:h-preprocessing}, which rescales pixel intensities; examples of $16\times 16$ GRD chips with their \texttt{is\_vessel} labels illustrate the visual similarity between classes. 
We then split the data into 80\% training and 20\% testing sets with stratified sampling, flatten the chips, and apply PCA to obtain lower-dimensional real (GRD) or complex (SLC) feature vectors.
\textbf{(b)} The PCA-transformed training vectors $\mathbf{x}_i$ are encoded using the Ry1DSt quantum kernel (though any quantum or classical kernel could be used; classical kernels require no data-encoding unitary and are computed using formulas such as Equations~\eqref{LinearKernel}, \eqref{RBFKernel}, or \eqref{LaplacianKernel}). The resulting training kernel matrix is passed to an SVC or KRC classifier, whose hyperparameters are selected via 10-fold cross-validation before retraining on the full training set to obtain the final trained model.
\textbf{(c)} The same encoding is applied to both training $\mathbf{x}_i$ and testing $x_i$ vectors to compute the testing kernel matrix, which is combined with the trained model to predict labels for the test samples. Performance metrics are then computed using the true and predicted labels.}
\label{fig:Workflow_schematic}
\end{figure*}

Each of the 6 datasets described in the previous subsections defines a binary classification problem, with input data samples given by the preprocessed vectors, and labels given by the associated \texttt{is\_vessel} or \texttt{is\_fishing} values.
A visual depiction of the machine learning approach that we employ can be seen in Figure~\ref{fig:Workflow_schematic}(b) and \ref{fig:Workflow_schematic}(c).
Specifically, we approach the learning tasks by seeking a classical or quantum kernel-based model which can accurately predict the \texttt{is\_vessel} or \texttt{is\_fishing} labels for the testing data.

The first step in the learning procedure involves feeding the preprocessed training vectors obtained via PCA to the data-encoding unitary that defines the quantum kernel we are working with (see Section~\ref{sec:Quantum kernel methods}). This data-enconding step is not required when using classical kernels, and we instead pass the preprocessed training vectors to a formula such as is Equations~\eqref{LinearKernel}, \eqref{RBFKernel} or \eqref{LaplacianKernel} to calculate the kernel matrix entries.

Next, once the entire training kernel matrix has been calculated, we pass the kernel matrix to a classical kernel-based ML algorithm, such as SVC or KRC (see Section~\ref{sec:Support vector classification} and \ref{sec:Kernel ridge classification}).
Using this algorithm, we perform a 10-fold cross-validation on the training data to find suitable hyperparameter values for the (classical or quantum) kernels using a grid search \cite{Mohri2018Foundations}. 
These hyperparameter values are listed in Table~\ref{tab:Hyperparameter_Values} of Appendix~\ref{sec:Hyperparameter values and configuration details}. 
Of the trialled values, the best value of each hyperparameter is selected based on the 
\emph{average} validation accuracy over all 10 folds of the training data. 
These hyperparameter values are then used to re-train a model on the complete training dataset.

Finally, after the training stage is complete, we feed the preprocessed vectors from both the training and testing datasets to the data-encoding unitary to calculate another kernel matrix for testing. 
This is equivalent to calculating the terms in the sum which defines the trained model in Equations~\eqref{SVCModel} and \eqref{KRCModel}. 
The prediced labels are compared to the true labels to assess performance.

\subsection{Kernels}
\label{sec:Kernels}

We now provide definitions of the various classical and quantum kernels which are trialled and used in the workflow described in the previous subsection (Figure~\ref{fig:Workflow_schematic}).
We start by defining the classical kernels.

The first classical kernel is the \emph{linear kernel}, denoted \(\mathcal{K}_{\textrm{lin}}:\mathcal{X}\times\mathcal{X}\to\R\), which is defined by the standard Euclidean inner product of its arguments,
\begin{equation}
\label{LinearKernel}
\mathcal{K}_{\textrm{lin}}(x,x^\prime) = \langle x,x^\prime\rangle.
\end{equation}
The second is the \emph{radial basis function} (RBF) \emph{kernel}, denoted \(\mathcal{K}_{\textrm{RBF}}:\mathcal{X}\times\mathcal{X}\to\R\), defined such that
\begin{equation}
\label{RBFKernel}
\mathcal{K}_{\textrm{RBF}}(x,x^\prime) = e^{-\gamma\|x-x^\prime\|^2},
\end{equation}
where \(\|\cdot\|\) is the standard Euclidean norm. 
The third and final classical kernel is the \emph{Laplacian kernel}, denoted \(\mathcal{K}_{\textrm{Lap}}:\mathcal{X}\times\mathcal{X}\to\R\), defined such that
\begin{equation}
\label{LaplacianKernel}
\mathcal{K}_{\textrm{Lap}}(x,x^\prime) = e^{-\gamma\|x-x^\prime\|_1},
\end{equation}
where \(\|\cdot\|_1\) is the \(l_1\) norm. 
We choose to trial these classical kernels since the RBF and Laplacian kernels are known to perform well compared with other standard kernels on a variety of problems~\cite{Tanner2025Learning,Virmani2023Comparative}, including in the context of SAR imagery~\cite{Yekkehkhany2014Comparison}, while the linear kernel provides an interpretable baseline.

We now define the quantum kernels trialled and simulated (without noise) in this work. 
As discussed in Section~\ref{sec:Quantum kernel methods}, specifying a quantum kernel amounts to choosing a data-encoding unitary acting on the initial state $\ket{0}$, where all qubits begin in the $+1$-eigenstate of the Pauli-Z operator. We consider three such unitaries: two for real-valued features and one for complex-valued features. Each unitary uses one qubit per feature, so trialling between 1 and 12 principal components corresponds to using between 1 and 12 qubits. For each architecture, we also vary the number of base layers (2, 3, or 4), excluding a single layer since it would produce no entanglement with the kernels considered.

We additionally tune the bandwidth hyperparameter~\cite{Canatar2023Bandwidth,Shaydulin2022Importance}, denoted $\beta\in\R$, which scales the input features prior to encoding and has been shown to substantially improve the generalisation performance of quantum kernel models. 
For example, in~\cite{Egginger2024Hyperparameter,Schnabel2025Scrutiny}, the authors found that tuning the bandwidth and regularisation for quantum kernels generally dominated the observed performance.
Thus, for each data sample $x\in\mathcal{X}$, we apply the transformation $x\mapsto\beta x$ before feeding it to the data-encoding unitary. 
Definitions of all quantum gates used here can be found in Sections 4.2 and 4.3 of~\cite{Nielsen2000Quantum}.

\begin{figure}[H]
    \includegraphics[width=0.4\textwidth]{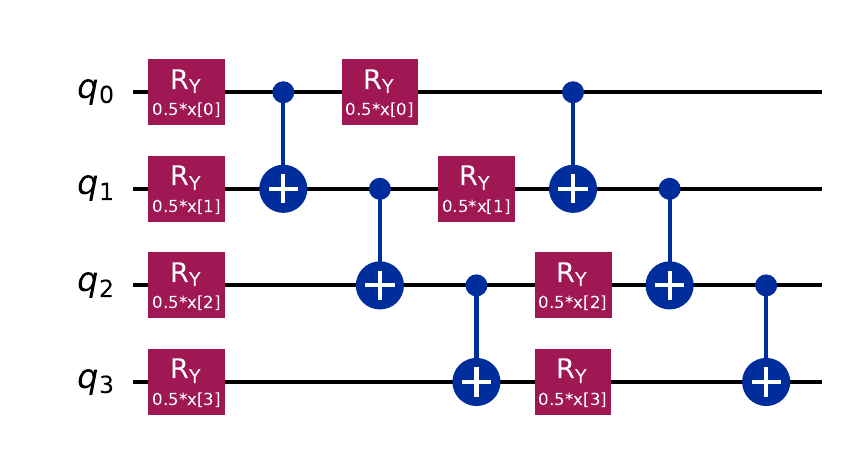}
    \caption{\footnotesize \justifying The data-encoding unitary for the Ry1DSt quantum kernel with 4 qubits, 2 layers and a bandwidth of $\beta=0.5$.
    Here \texttt{x[i]} represents the $i^{\textrm{th}}$ component of the input $x\in\R^4$.}
    \label{fig:Ry_1DStaircaseCxKernel}
\end{figure}

The first quantum kernel, called the \emph{Ry1DSt kernel}, has a base layer that includes $R_Y$ rotation gates on each qubit, rotating the states of the qubits by an angle proportional to the associated real input feature. 
Following the $R_Y$ gates in each base layer is a 1-dimensional (1D) sequence of CNOT gates arranged in a ``staircase'' pattern (thus we include ``St'' in the name of this kernel).
An example of the data-encoding unitary can be found in Figure~\ref{fig:Ry_1DStaircaseCxKernel} with 4 qubits, 2 layers and a bandwidth of $\beta=0.5$.

\begin{figure}[H]
    \includegraphics[width=0.4\textwidth]{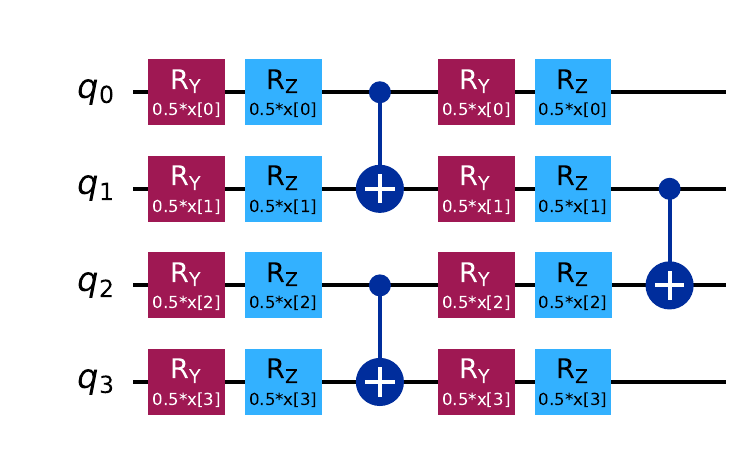}
    \caption{\footnotesize \justifying The data-encoding unitary for the RyRz1DAlt quantum kernel with 4 qubits, 2 layers and a bandwidth of $\beta=0.5$.
    Here \texttt{x[i]} represents the $i^{\textrm{th}}$ component of the input $x\in\R^4$.}
    \label{fig:RyRz_1DAlternatingCxKernel}
\end{figure}

The second quantum kernel, called the \emph{RyRz1DAlt kernel}, has a base layer that includes $R_Y$ and $R_Z$ rotation gates on each qubit which both rotate by an angle proportional to the associated real input feature. 
Following the rotation gates is a collection of CNOT gates. In one layer the CNOT gates are controlled on even numbered qubits and target the qubit one position below. In the next layer the CNOT gates are instead controlled on odd numbered qubits. Since we alternate between these two layers we include ``Alt'' in the name of this kernel.
An example of the data-encoding unitary can be found in Figure~\ref{fig:RyRz_1DAlternatingCxKernel} with 4 qubits, 2 layers and a bandwidth of $\beta=0.5$.

The third and final quantum kernel, called the \emph{CRyRz1DSt kernel}, has a base layer that includes $R_Y$ and $R_Z$ rotation gates. The angle for the  $R_Y$ gate is proportional to the modulus of the associated complex input feature, while the angle for the $R_Z$ gate is proportional to the phase.
We include ``C'' at the beginning of the name of this kernel to indicate that the kernel takes complex arguments as input.
Following the rotation gates is a collection of CNOT gates arranged again in a ``staircase'' pattern, identical to the arrangement of CNOT gates in the RY1DSt kernel. 
An example of the data-encoding unitary can be found in Figure~\ref{fig:Complex_RyRz_1DStaircaseCxKernel} with 4 qubits, 2 layers and a bandwidth of $\beta=0.5$.

\begin{figure}[H]
    \includegraphics[width=0.5\textwidth,trim={1.35cm 0cm 0.75cm 0cm}, clip]{Complex_RyRz_1DStaircaseCx.pdf}
    \caption{\footnotesize \justifying The data-encoding unitary for the CRyRz1DSt quantum kernel with 4 qubits, 2 layers and a bandwidth of $\beta=0.5$.
    Here \texttt{abs(x[i])} and \texttt{arg(x[i])} represent the magnitude and phase (respectively) of the $i^{\textrm{th}}$ component of the input $x\in\C^4$.}
    \label{fig:Complex_RyRz_1DStaircaseCxKernel}
\end{figure}

The Ry1DSt unitary operator is also sometimes called a hardware efficient ansatz~\cite{Kandala2017HEA,Leone2024HEA}. 
A close variant of this unitary was used to study exponential concentration in~\cite{Thanasilp2024Exponential}, though the analysis in~\cite{Thanasilp2024Exponential} did not consider the bandwidth parameter which may help to remedy the exponential concentration phenomena~\cite{Shaydulin2022Importance}.
Either way, the appearance of this circuit in these works motivates our consideration of the circuit here.
The RyRz1DAlt circuit was similarly employed in kernel-based learning in~\cite{Haug2023Quantum} and adopted in~\cite{Schnabel2025Scrutiny}, which motivates our inclusion of the circuit in the current work.
Note that a comprehensive ablation study would be valuable, but is beyond the scope of this work. 
We therefore restrict attention to a representative subset of kernels previously considered in the literature, enabling a focused investigation within a manageable experimental setting.

Contrarily, we are not aware of the CRyRz1DSt kernel being used in prior works, but we employ this kernel here in an attempt to meaningfully encode phase information from the preprocessed vector representations of the SLC chips.
The justification for this is that the associated data-encoding unitary injects a relative phase between the components of the state of each qubit (in the eigenbasis of the Pauli-$Z$ operator) equal to the complex argument of the associated complex input feature.
Also notice that by scaling the input features by the real bandwidth $\beta\in\R$, we preserve the complex phases of the input features.

\subsection{Implementation}

In this section we provide details about the computational implementation of the machine learning workflow depicted in Figure~\ref{fig:Workflow_schematic}.
Firstly, in the preprocessing stage, the application of the function $h$ is straightforward and was implemented with a custom function, while the stratified sampling of the datasets into training and testing datasets was implemented with the \texttt{sklearn}~\cite{Pedregosa2011Scikit} function \texttt{train\_test\_split}.
Next, to apply PCA we implement a standard \texttt{ComplexPCA} class since \texttt{sklearn} does not support complex data. 
To check the correctness of our implementation we tested our class against \texttt{sklearn.PCA} using real GRD inputs, and observed the results to be identical for the GRD data.

In the training and testing stages, the kernel matrices derived from a quantum kernel were calculated using the \texttt{qiskit}~\cite{Javadi2024Qiskit} function \texttt{FidelityQuantumKernel}. 
The simulations of the quantum kernels were performed without noise. 
A full investigation of the impact of noise in this context is beyond the scope of this paper, but for discussions of the impact of noise on QKMs, we refer the reader to~\cite{Wang2021Towards,Heyraud2022NoisyQKMs}.

Following the kernel calculations, we performed the 10-fold cross-validation by grid search with the \texttt{sklearn} class \texttt{model\_selection.GridSearchCV} with \texttt{svm.SVC} and a custom KRC estimator based on \texttt{kernel\_ridge.KernelRidge} to implement SVC and KRC (respectively).
These methods automatically calculate the kernel matrices for various standard classical kernels, while the quantum kernel matrices were passed as \texttt{precomputed} kernels.
The learning performance metrics are then calculated using the \texttt{sklearn} function \texttt{classification\_report}.

\section{Results and Discussion}
\label{sec:Results and discussion}

In this section we present and discuss the results obtained with each kernel applied to each dataset for both the \texttt{is\_vessel} and \texttt{is\_fishing} classification tasks.
In Table~\ref{tab:Results_Summary}, we provide the best learning performance metrics\footnote{The definitions of these metrics can be found in Section IV.A of~\cite{Guan2023Fishing}}—accuracy, precision, recall, and $F_1$-score—obtained on both the training and testing sets with the kernel-based models.
The models were trained on the full training dataset using hyperparameters tuned via a 10-fold cross-validation on the same dataset.
Note that the reported precision, recall, and $F_1$-scores are for the \texttt{true} class in each task.

In Appendix~\ref{sec:Hyperparameter values and configuration details}, Tables~\ref{tab:Hyperparameter_Values} and \ref{tab:Results_Hyperparameters} provide all trialled hyperparameter values and configuration details used to obtain the results.
In Appendix~\ref{sec:Principal component analysis considerations} we provide plots of the results, and the cumulative explained variance ratio for the GRD datasets, as functions of the number of PCA components.
Further results, including the precision, recall, and $F_1$-scores obtained on the \texttt{false} class of the training and testing datasets, together with the macro and weighted averages of the reported metrics, can be found in the Github repository (see Data Availability).

\begin{table*}
\begin{adjustbox}{center,max width=\textwidth}
\begin{tabular}{ | c | c | c | c | c | c | c | c | c | c | c |}
\cline{1-11}
\multirow{2}{*}{\textbf{Task}}
& \multirow{2}{*}{\textbf{Data type}} 
& \multirow{2}{*}{\textbf{Kernel}} 
& \multicolumn{2}{ c |}{\textbf{Accuracy}} 
& \multicolumn{2}{ c |}{\textbf{Precision}} 
& \multicolumn{2}{ c |}{\textbf{Recall}} 
& \multicolumn{2}{ c |}{\textbf{$F_1$ score}} \\
\cline{4-11}
& & &
Training & Testing &
Training & Testing &
Training & Testing &
Training & Testing 
\\
\cline{1-11}
\multirow{7}{*}{\textbf{is\_vessel}}
    & \multirow{5}{*}{$16\times16$ GRD} 
        & Linear 
            & 0.7060 & 0.7400 & 0.7036 & 0.7459 & 0.7120 & 0.7280 & 0.7078 & 0.7368 \\
    &
        & Laplacian 
            & \textbf{1.0000} & 0.8840 & \textbf{1.0000} & \textbf{0.9000} & \textbf{1.0000} & 0.8640 & \textbf{1.0000} & 0.8816 \\
    &
        & RBF 
            & 0.8970 & 0.8560 & 0.9214 & 0.8739 & 0.8680 & 0.8320 & 0.8939 & 0.8525 \\
    &
        & Ry1DSt 
            & 0.9670 & \textbf{0.8920} & 0.9698 & 0.8889 & 0.9640 & \textbf{0.8960} & 0.9669 & \textbf{0.8924} \\
    &
        & RyRz1DAlt 
            & 0.9450 & 0.8760 & 0.9423 & 0.8917 & 0.9480 & 0.8560 & 0.9452 & 0.8735 \\
\cline{2-11}
    & {$16\times16$} SLC 
        & CRyRz1DSt 
            & \textbf{1.0000} & 0.6640 & \textbf{1.0000} & 0.7356 & \textbf{1.0000} & 0.5120 & \textbf{1.0000} & 0.6038 \\
\cline{2-11}
    & {$70\times12$} SLC 
        & CRyRz1DSt 
            & \textbf{1.0000} & 0.6560 & \textbf{1.0000} & 0.6612 & \textbf{1.0000} & 0.6400 & \textbf{1.0000} & 0.6504 \\
\cline{1-11}
\multirow{7}{*}{\textbf{is\_fishing}}
    & \multirow{5}{*}{$16\times16$ GRD} 
        & Linear
            & 0.7980 & 0.7920 & 0.7411 & 0.7386 & 0.9160 & 0.9040 & 0.8193 & 0.8129 \\
    & 
        & Laplacian
            &  0.9990 & \textbf{0.9040} & 0.9980 & 0.8915 & \textbf{1.0000} & 0.9200 & 0.9990 & \textbf{0.9055} \\
    &
        & RBF 
            & 0.9960 & 0.8800 & 0.9921 & 0.8417 & \textbf{1.0000} & \textbf{0.9360} & 0.9960 & 0.8864 \\
    &
        & Ry1DSt 
            & 0.9800 & 0.8920 & 0.9615 & 0.8603 & \textbf{1.0000} & \textbf{0.9360} & 0.9804 & 0.8966 \\
    &
        & RyRz1DAlt 
            & 0.9990 & \textbf{0.9040} & 0.9980 & \textbf{0.9391} & \textbf{1.0000} & 0.8640 & 0.9990 & 0.9000 \\
\cline{2-11}
    & {$16\times16$} SLC 
        & CRyRz1DSt 
            & \textbf{1.0000} & 0.8000 & \textbf{1.0000} & 0.7586 & \textbf{1.0000} & 0.8800 & \textbf{1.0000} & 0.8148 \\
\cline{2-11}
    & {$70\times12$} SLC 
        & CRyRz1DSt 
            & \textbf{1.0000} & 0.7920 & \textbf{1.0000} & 0.8349 & \textbf{1.0000} & 0.7280 & \textbf{1.0000} & 0.7778 \\
\cline{1-11}
\end{tabular}
\end{adjustbox}
\caption{\footnotesize \justifying The best learning performance metrics obtained by the models trained on the full training dataset for the \texttt{is\_vessel} and \texttt{is\_fishing} classification tasks using each kernel with the hyperparameter values listed in Table~\ref{tab:Results_Hyperparameters}.
The best value of each reported metric obtained for either task on the training and testing datasets are highlighted in bold.}
\label{tab:Results_Summary}
\end{table*}

\subsection{Results for the \texttt{is\_vessel} task}

As shown in Table~\ref{tab:Results_Summary}, the Ry1DSt kernel achieved the highest accuracy of 0.8920, the highest recall of 0.8960, and the highest $F_1$-score of 0.8924 on the testing dataset for the \texttt{is\_vessel} classification task. 
On the other hand, the Laplacian kernel obtained the highest testing precision of 0.9000, but exhibits perfect training scores, suggestive of overfitting.
Note that all these results were obtained with the GRD data.

In the case of the complex SLC data, the CRyRz1DSt kernel achieves perfect training performance.
However, it generalises poorly, with test accuracies between 0.65 and 0.67, far below the linear kernel which obtained a test accuracy of 0.7400 with the GRD chips. 
This indicates significant overfitting, more severe than for the Laplacian kernel. 
As shown in Figure~\ref{fig:Accuracies_vs_Num_PCA_Components} of Appendix~\ref{sec:Principal component analysis considerations}, training performance becomes near-perfect when more than 5 principal components are used. 
Despite its design, the results indicate that the CRyRz1DSt kernel does not effectively exploit phase information, possibly due to a mismatch with the task or the effects of PCA preprocessing.

Overall, based on test performance, the best performing kernel for the \texttt{is\_vessel} classification task was the Ry1DSt kernel, followed closely by the Laplacian kernel.

\subsection{Results for the \texttt{is\_fishing} task}

Table~\ref{tab:Results_Summary} shows that using the GRD data, both the RyRz1DAlt and Laplacian kernel obtained the greatest test accuracy of 0.9040 for the \texttt{is\_fishing} task. 
The RyRz1DAlt kernel also obtained the greatest test precision of 0.9391, while the Laplacian obtained the greatest test $F_1$-score of 0.9055.
Meanwhile, the greatest test recall of 0.9360 was obtained by both the Ry1DSt and RBF kernels.

When examining performance on the complex SLC data, we see the same trend as for the \texttt{is\_vessel} task.
The CRyRz1DSt kernel obtains perfect training scores with upwards of 5 principal components (Appendix~\ref{sec:Principal component analysis considerations}), but less than ideal testing scores, albeit with better performance here than on the \texttt{is\_vessel} task. 
The testing scores in this case are similar or just marginally better than those obtained with the linear kernel, suggesting again that the CRyRz1DSt kernel may not be effectively exploiting the phase information in the PCA preprocessed SLC chips.

In the case of the \texttt{is\_fishing} classification task, the best performing kernels in terms of testing performance were the RyRz1DAlt and Laplacian kernels.

\subsection{Insights and Limitations}

The results show that quantum kernels, especially the Ry1DSt and RyRz1DAlt kernels, perform competitively with the strong baselines obtained with the Laplacian and RBF kernels.
Both of these quantum kernels obtained higher testing metric values than the RBF kernel in all cases except in the \texttt{is\_fishing} task, where the RBF and Ry1DSt kernels obtained equally higher test recall than the RyRz1DAlt kernel.
This suggests that quantum machine learning methods, specifically QKMs, have the potential to provide enhanced learning performance for the purpose of improving IUU fishing surveillance capabilities.

However, these results are based on classical simulations. 
Implementing the proposed kernels on quantum hardware would likely introduce additional challenges~\cite{Thanasilp2024Exponential,Heyraud2022NoisyQKMs,Wang2021Towards}. 
Furthermore, the analysis in Appendix~\ref{sec:Principal component analysis considerations} indicates diminishing returns when increasing the number of principal components (and hence qubits). 
While this does not strongly support scaling up qubit counts, the results nevertheless provide evidence that QKMs can achieve competitive performance.
Taken together, these findings support the promise of QKMs, while highlighting the need for further research to justify experimental implementations.

The reasonably good performance of the linear kernel, which obtained accuracies of 0.7400 and 0.7920 on the \texttt{is\_vessel} and \texttt{is\_fishing} tasks (respectively), suggests that the GRD datasets have a moderate degree of linear separability. 
In Appendix~\ref{sec:t-SNE plots of the GRD datasets}, we provide two-dimensional visualisations (obtained using $t$-distributed stochastic neighbor embeddings) of the GRD datasets for the \texttt{is\_vessel} and \texttt{is\_fishing} classification tasks.
The plots show that the GRD \texttt{is\_fishing} dataset exhibits a greater degree of linear separability than the GRD \texttt{is\_vessel} dataset, consistent with the performance of the linear kernel on these datasets.
However, to improve performance further, the results show that the employment of non-linear kernels is necessary, which motivates their application in this context.

The accuracies obtained in this work are competitive with those reported in the related works discussed in Section~\ref{sec:Related Work}, with only a few exceptions being the test accuracies obtained in~\cite{Wang2018Ship,Ji2013Ship,Miller2023Quantum}.
However each of these works focused on different datasets, and the first two did not incorporate any quantum aspects in their models, which makes a direct comparison hard to justify.
In contrast, one of the related works which dealt with the SARFish dataset is~\cite{Williams2023Contrastive}, in which the authors reported obtaining $F_1$ scores of 0.871, 0.871 and 0.866 by utilising $32\times32$ GRD, $23\times140$ SLC and the magnitude of $140\times23$ SLC chips. 
Our work differs from this reference since the authors tackled the classification task as a 3-class problem, but our best $F_1$ scores of 0.8924 and 0.9055 for the two tasks are higher.

Another relevant work for comparison is~\cite{Paolo2022xView3}, which describes the xView3 challenge involving a subset of the SARFish dataset.
The winner of the xView3 challenge, BloodAxe, reported $F_1$-scores on the \texttt{is\_vessel} and \texttt{is\_fishing} classification tasks of 0.9392 and 0.8425 (respectively).
Once again, comparing with the $F_1$-scores obtained in this work, we see that we achieved a greater $F_1$-score for the latter task using the Laplacian, RBF, Ry1DSt, and RyRz1DAlt kernels, but worse $F_1$-scores for the former task.
However it should be noted that BloodAxe's machine learning workflow not only dealt with the classification aspect, but also maritime object detection and vessel length regression. 

\section{Conclusion and Future Work}
\label{sec:Conclusion and Future Work}

In this work, we investigate the efficacy of quantum kernel methods in maritime object classification using SAR imagery from the Sentinel-1 sensor for the purposes of IUU fishing surveillance.
Specifically, using both real GRD and complex SLC chip images, we compare the performance of a representative collection of quantum kernels with classical linear, Laplacian, and RBF kernels.
We restrict the comparison to be between just kernel based models so that the comparison is as fair and meaningful as possible.
The results show that quantum kernels can obtain competitive learning performance metric values, either matching or exceeding those obtained with the classical kernels.
This suggests that quantum kernels may provide opportunities for quantum-enhanced learning, supporting efforts to combat IUU fishing.

However, performance is not uniform across all settings. 
The CRyRz1DSt kernel applied to the complex SLC data does not perform competitively and exhibits strong overfitting. 
This suggests that, under the PCA-based preprocessing pipeline, the kernel is not effectively exploiting the phase information present in the complex SLC data.
More generally, the results indicate diminishing returns when increasing the number of principal components (and hence qubits), including in the GRD-based experiments. 
This suggests that simply increasing the feature dimension under the current encoding scheme is unlikely to yield substantial improvements in classification performance.
Overall, the results demonstrate that competitive performance with quantum kernel methods is achievable in this setting, but highlight the importance of encoding and preprocessing choices.

The work naturally lends itself to several possible directions for future research.
First, despite numerical evidence to the contrary, improved performance achieved by increasing the number of qubits used in the quantum kernels cannot be ruled out. 
For example, the quantum kernels applied to the GRD \texttt{is\_vessel} datasets performed best with 12 qubits, suggesting that further exploration of higher qubit counts may be worthwhile.
However, we could not extend our analysis further in this direction due to computational constraints.
In addition, evaluating the performance of these models on real quantum hardware would provide valuable insight into their practical robustness.

From another perspective, it would be interesting to systematically reduce the amount of training data supplied to the models. 
This would allow us to investigate whether there are differences in the generalisation capabilities of classical and quantum kernels when only small datasets are available. 
Changing the amount of training data might also help to close the generalisation gap exhibited by the CRyRz1DSt kernel.

Similarly, exploring alternative encoding and preprocessing methods appears important to further improve performance and enhance learning outcomes.
Fault-tolerant quantum hardware may enable scaling to substantially larger qubit counts with more sturcturally appropriate encodings than are accessible via classical simulation.
For example, one may find value in investigating whether learning performance can be improved by using qubits connected on a two-dimensional lattice, to which an image can be passed in its raw form, with one pixel encoded per qubit.

Finally, our results highlight the need for methods that more effectively exploit the phase information present in complex SAR data. 
Developing improved preprocessing and encoding strategies for complex-valued inputs within quantum kernel frameworks represents a promising and relatively unexplored direction for future research.
Despite the negative result obtained with the CRyRz1DSt kernel, the fact that quantum models operate natively in complex Hilbert spaces provides a strong motivation to further investigate their suitability in this context.

\section*{Acknowledgements}

This project is led by the Centre for Quantum Information, Simulation and Algorithms (QUISA) and supported by the Defence Science and Technology Group (DSTG) and the Advanced Strategic Capabilities Accelerator (ASCA) through its Emerging and Disruptive Technologies (EDT) Program. Substantial computational resources were provided by the Pawsey Supercomputing Research Centre. The authors thank Jack Blyth and Tuan Nguyen for their insightful discussions and suggestions on the machine learning and preprocessing methods used in this work. J.T. acknowledges receiving an Australian Government Research Training Program (RTP) Scholarship.

\vskip0.25cm
\(\)

\section*{Data Availability}
\label{sec:Data Availability}

The datasets analysed in this study are derived from the publicly available SARFish dataset, as described in \cite{Luckett2024Sarfish}. 
Links to download the original dataset are provided therein. 
We provide all code necessary to extract the GRD and SLC chips from the SARFish dataset and to reproduce the exact training and testing splits used in our experiments. 
These resources are available in our public GitHub repository at \href{https://github.com/John-J-Tanner/Extract-SARFish-Data}{https://github.com/John-J-Tanner/Extract-SARFish-Data}.

\bibliographystyle{unsrtnat}

\bibliography{BibliographyV2}

\clearpage

\appendix

\onecolumngrid

\section{Hyperparameter values and configuration details}
\label{sec:Hyperparameter values and configuration details}

In this appendix, Table \ref{tab:Hyperparameter_Values} provides a list of the hyperparameter values which were trialled for each kernel during the 10-fold cross-validation performed during the training procedure.
Table~\ref{tab:Results_Hyperparameters} then specifies the algorithm and hyperparameter values that were used to obtain the learning performance metrics report in Table~\ref{tab:Results_Summary} of Section \ref{sec:Results and discussion}.

\begin{table}[H]
\begin{adjustbox}{max width=\textwidth,center}
\begin{tabular}{| c | c | c |}
\cline{1-3}
\multicolumn{1}{|c|}{\textbf{Kernel(s)}}& \multicolumn{1}{c|}{\textbf{Hyperparameter}}
& \multicolumn{1}{c|}{\textbf{Cross-validated hyperparameter values}}\\
\cline{1-3}


\multirow{5}{*}{All kernels}
& \multirow{2}{*}{\(C\)} & \footnotesize\(\{\,10^{-4},\,10^{-3.5},\,10^{-3},\,10^{-2.5},\,10^{-2},\,10^{-1.5},\,10^{-1},\,10^{-0.5},\)\\
& & \footnotesize\(\,1,\,10^{0.5},\,10,\,10^{1.5},\,10^2,\,10^{2.5},\,10^3,\,10^{3.5},\,10^4\,\}\)\\
& \multirow{2}{*}{\(\lambda\)} & \footnotesize\(\{\,10^{-4},\,10^{-3.5},\,10^{-3},\,10^{-2.5},\,10^{-2},\,10^{-1.5},\,10^{-1},\,10^{-0.5},\)\\
& & \footnotesize\(\,1,\,10^{0.5},\,10,\,10^{1.5},\,10^2,\,10^{2.5},\,10^3,\,10^{3.5},\,10^4\,\}\)\\
& \multirow{1}{*}{PCA components} & \footnotesize\(\{\,1,\,2,\,3,\,4,\,5,\,6,\,7,\,8,\,9,\,10,\,11,\,12\,\}\)\\
\cline{1-3}


\multirow{2}{*}{RBF, Laplacian}
& \multirow{2}{*}{\(\gamma\)} & \footnotesize\(\{\,10^{-4},\,10^{-3.5},\,10^{-3},\,10^{-2.5},\,10^{-2},\,10^{-1.5},\,10^{-1},\,10^{-0.5},\)\\
 & & \footnotesize\(\,1,\,10^{0.5},\,10,\,10^{1.5},\,10^2,\,10^{2.5},\,10^3,\,10^{3.5},\,10^4\,\}\)\\
\cline{1-3}


\multirow{2}{*}{Ry1DSt, RyRz1DAlt, CRyRz1DSt}
& \(\beta\) &\footnotesize\(\{\,0.1,\,0.2,\,0.3,\,0.4,\,0.5,\,0.6,\,0.7,\,0.8,\,0.9,\,1.0\,\}\)\\
& \multirow{1}{*}{Layers} & \footnotesize\(\{2,3,4\}\)\\
\cline{1-3}

\end{tabular}
\end{adjustbox}
\caption{Lists of hyperparameters values which were trialled for each kernel during the 10-fold cross-validation performed on the training datasets. 
The hyperparameters $C$ and $\lambda$ refer to the regularisation strength parameters in Equations~\eqref{SVC} and \eqref{KRC}. 
The number of principal components (PCA components) corresponds to the dimensions of the preprocessed vectors obtained via PCA (see Figure~\ref{fig:Workflow_schematic}).
In the case of all quantum kernels, the number of principal components also equates to the number of qubits in the associated data-encoding unitary.
The hyperparameter $\gamma$ corresponds to the parameter appearing in definitions of the RBF and Laplacian kernels in Equations~\eqref{RBFKernel} and \eqref{LaplacianKernel}.
The hyperparameter $\beta$ corresponds to the quantum bandwidth hyperparameter discussed in the fourth paragraph of Section~\ref{sec:Kernels}.
The number of layers (Layers) correspond to the number of times that the base layer of each quantum kernel is repeated in the associated data-encoding unitary.}
\label{tab:Hyperparameter_Values}
\end{table}

\begin{table}[H]
\begin{adjustbox}{max width=\textwidth,center}
\begin{tabular}{| c | c | c | c | c | c | c | c | c |}

\cline{1-9}
\textbf{Task} & \textbf{Dataset} & \textbf{Kernel} & \textbf{Algorithm} & \textbf{PCA components} & \textbf{$C$ (SVC) or $\lambda$ (KRC)} & $\gamma$ & $\beta$ & \textbf{Layers} \\

\cline{1-9}

\multirow{7}{*}{\texttt{is\_vessel}} & \multirow{5}{*}{$16\times16$ GRD} 
& Linear & KRC & 10 & $10^{-4}$ & $\times$ & $\times$ & $\times$ \\

&& Laplacian & KRC & 9 & $10^{-1}$ & $10^{-1}$ & $\times$ & $\times$ \\

&& RBF & SVC & 12 & 1 & $10^{-1.5}$ & $\times$ & $\times$ \\

&& Ry1DSt & SVC & 12 & $10^{0.5}$ & $\times$ & 0.3 & 3 \\

&& RyRz1DAlt & KRC & 12 & 1 & $\times$ & 0.2 & 3 \\

\cline{2-9}

&\multirow{1}{*}{$16\times16$ SLC} 
& CRyRz1DSt & SVC & 9 & $10^{-4}$ & $\times$ & 0.4 & 4 \\

\cline{2-9}

&\multirow{1}{*}{$70\times12$ SLC} 
& CRyRz1DSt& SVC & 10 & $10^{-4}$ & $\times$ & 0.5 & 3 \\

\cline{1-9}

\multirow{7}{*}{\texttt{is\_fishing}} & \multirow{5}{*}{$16\times16$ GRD} 
& Linear & SVC & 6 & $10^{-2.5}$ & $\times$ & $\times$ & $\times$ \\

&& Laplacian & KRC & 12 & $10^{-0.5}$ & $10^{-1}$ & $\times$ & $\times$ \\

&& RBF & KRC & 12 & 1 & $10^{-0.5}$ & $\times$ & $\times$ \\

&& Ry1DSt & KRC & 11 & $10^{-0.5}$ & $\times$ & 0.3 & 4 \\

&& RyRz1DAlt & SVC & 11 & 1 & $\times$ & 0.8 & 3 \\

\cline{2-9}

&\multirow{1}{*}{$16\times16$ SLC} 
& CRyRz1DSt & KRC & 11 & $10^{-0.5}$ & $\times$ & 0.2 & 2 \\

\cline{2-9}

&\multirow{1}{*}{$70\times12$ SLC} 
& CRyRz1DSt& KRC & 12 & 1 & $\times$ & 0.5 & 4 \\

\cline{1-9}

\end{tabular}
\end{adjustbox}
\caption{Algorithms (either SVC as discussed in Section~\ref{sec:Support vector classification} or KRC as discussed in Section~\ref{sec:Kernel ridge classification}) and hyperparameters used to obtain the learning performance metrics reported in Table~\ref{tab:Results_Summary}.
Note that when a cell contains $\times$, this indicates that the associated hyperparameter is not relevant to the kernel described by the associated row.}
\label{tab:Results_Hyperparameters}
\end{table}

\section{Principal component analysis considerations}
\label{sec:Principal component analysis considerations}

\begin{figure}[H]
    \centering
    \begin{tikzpicture}
        \node at (0,0) {\includegraphics[width=0.9\textwidth,trim={0cm 0cm 0cm 1.1cm},clip]{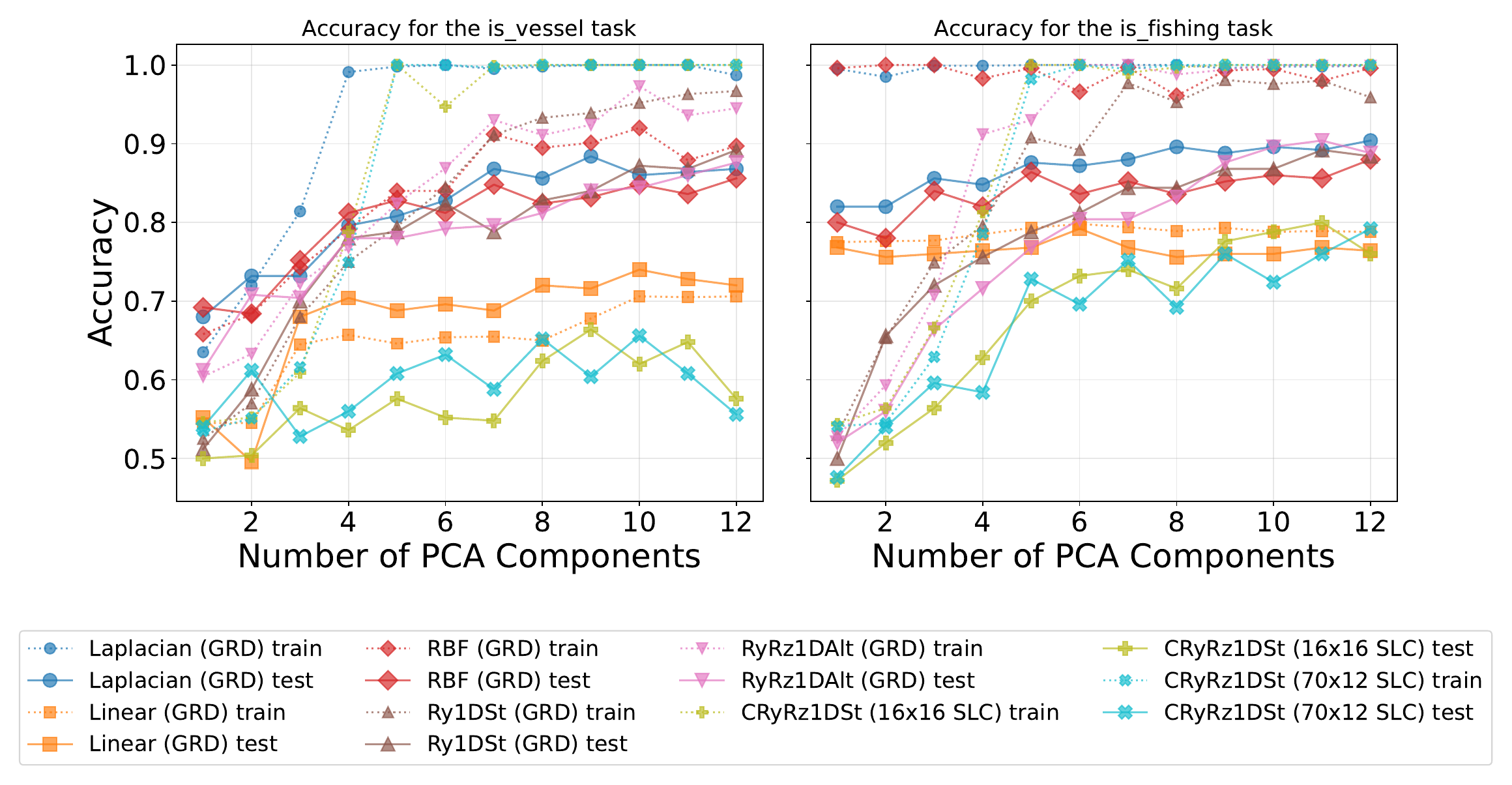}};
        \node at (-2.875,4.25) {\large\textbf{(a)}\texttt{ is\_vessel}};
        \node at (3.75,4.25) {\large\textbf{(b)}\texttt{ is\_fishing}};
    \end{tikzpicture}
    \caption{Accuracies obtained on the training and testing datasets with each of the models for the \textbf{(a)} \texttt{is\_vessel} and \textbf{(b)} \texttt{is\_fishing} classification tasks. Specifically, for each number of principal components (PCA components), we plot the accuracies obtained with the models trained on the full training dataset with the best hyperparameters trialled for that number of PCA components.}
    \label{fig:Accuracies_vs_Num_PCA_Components}
\end{figure}

In this appendix, Figure~\ref{fig:Accuracies_vs_Num_PCA_Components} shows the best accuracies achieved by all models on the \texttt{is\_vessel} and \texttt{is\_fishing} tasks as a function of the number of principal components. 
For the quantum kernels, we fix the algorithm, quantum bandwidth $\beta$, and number of layers as given in Table~\ref{tab:Results_Hyperparameters}, and tune only the regularisation via cross-validation for each number of principal components. 
For the classical kernels, all hyperparameters are tuned by cross-validation. 
In all cases, we report training and testing performance using the best trialled hyperparameter settings.
The results show that test accuracy generally improves as the number of principal components increases from 1, but plateaus at around 8 components. 
Notably, the CRyRz1DSt kernel achieves near-perfect training accuracy for more than 5 components, but exhibits poor generalisation, with test accuracy not exceeding 0.7 and 0.8 on the \texttt{is\_vessel} and \texttt{is\_fishing} tasks (respectively).

\begin{figure}[H]
    \centering
    \includegraphics[width=0.5\textwidth]{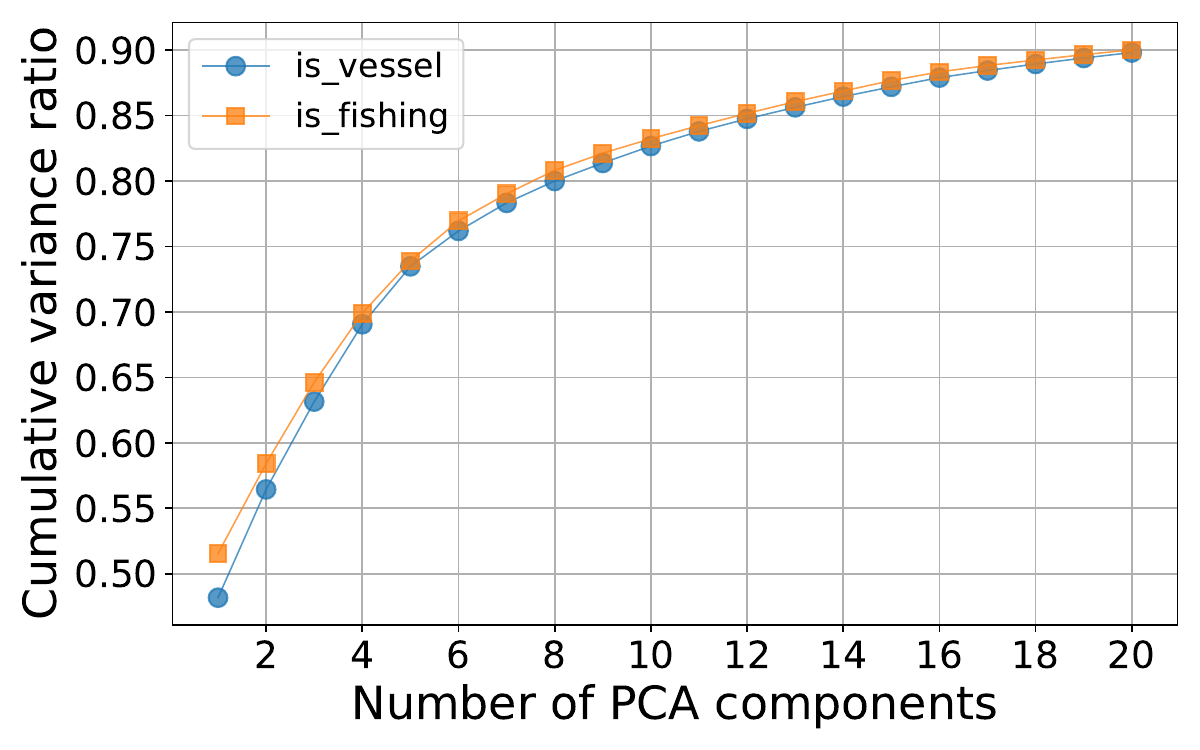}
    \caption{A plot of the cumulative explained variance ratio as a function of the number of principal components for the $16\times16$ GRD chips used for the \texttt{is\_vessel} and \texttt{is\_fishing} classification tasks.}
    \label{fig:PCA_Explained_Variance_Ratio}
\end{figure}

Figure~\ref{fig:PCA_Explained_Variance_Ratio} shows a plot of the cumulative explained variance ratios, as a function of the number of principal components, for the $16\times16$ GRD chips used in the \texttt{is\_vessel} and \texttt{is\_fishing} classification tasks. 
This figure shows that roughly $75\%$ of the variance in the GRD dataset is described by the first 5 principal components, with the following $15\%$ described by the next 15 components. 
This, together with Figure~\ref{fig:Accuracies_vs_Num_PCA_Components}, appears to suggest that there are diminishing returns associated with including more principal components in the data, especially when one considers that each extra feature requires another qubit to encode using the quantum kernels considered in this work.

\section{$t$-SNE plots of the GRD datasets}
\label{sec:t-SNE plots of the GRD datasets}

\begin{figure}[H]
    \centering
    \begin{tikzpicture}
    \node at (0,0) {\includegraphics[width=0.75\textwidth,trim={0cm 0cm 0cm 1.1cm},clip]{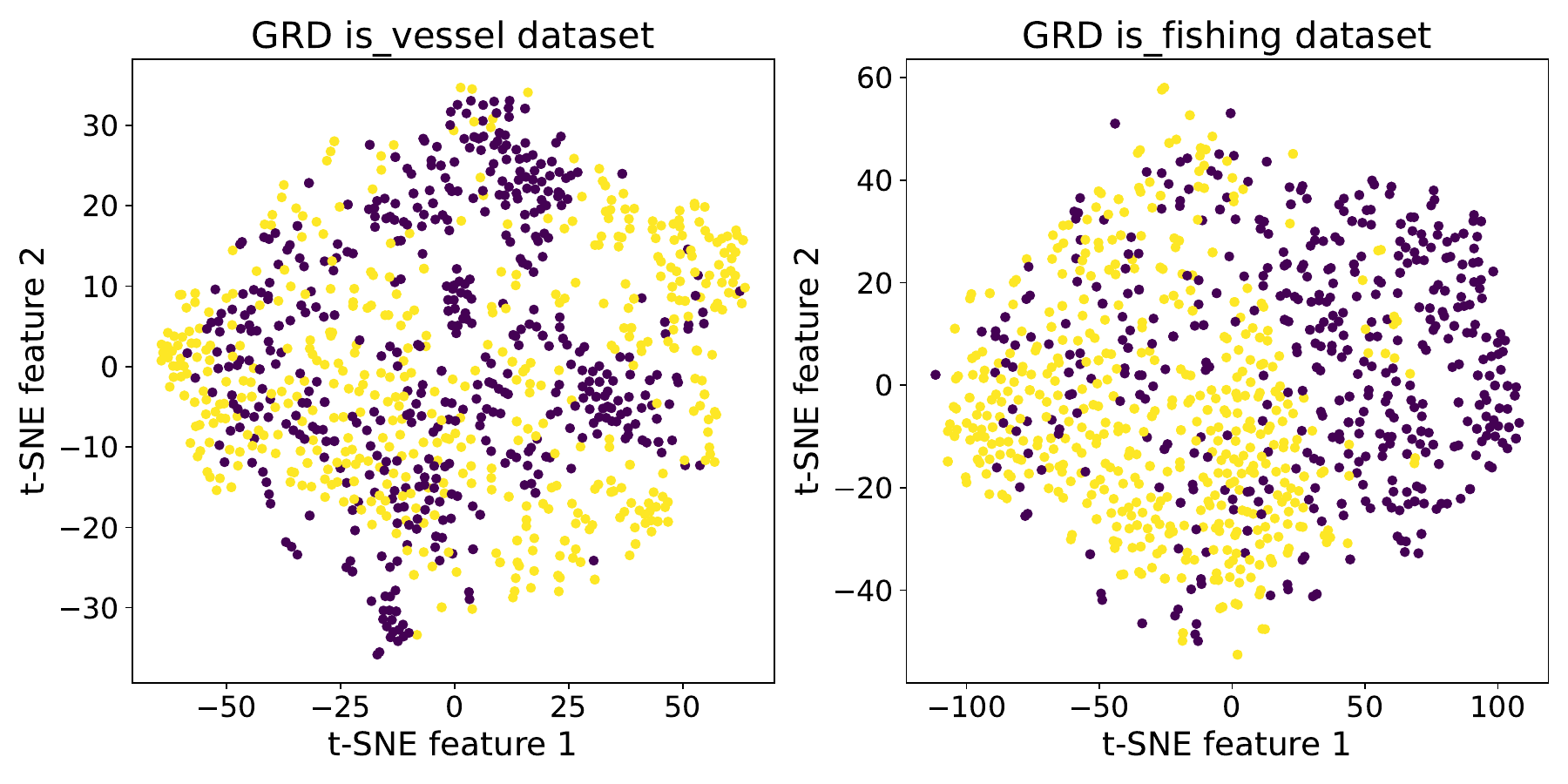}};
    \node at (-2.75,3.375) {\large\textbf{(a)} GRD \texttt{is\_vessel}};
    \node at (3.75,3.375) {\large\textbf{(b)} GRD \texttt{is\_fishing}};
    \end{tikzpicture}
    \caption{A two-dimensional visualisation of the GRD datasets used for the \textbf{(a)} \texttt{is\_vessel} and \textbf{(b)} \texttt{is\_fishing} classification tasks obtained using $t$-SNE with a perplexity of 30.}
    \label{fig:t-SNE}
\end{figure}

In this appendix, Figure~\ref{fig:t-SNE} shows a two-dimensional visualisation of the GRD datasets used for the \texttt{is\_vessel} and \texttt{is\_fishing} classification tasks.
The visualisations are obtained using $t$-distributed stochastic neighbor embedding ($t$-SNE).
The figure provides visual insight into the linear separability of the dataset, showing that the \texttt{is\_fishing} GRD dataset exhibits a reasonable degree of linear separability, while the \texttt{is\_vessel} GRD dataset appears less linearly separable.

\end{document}